\documentclass[onecollarge,numbook,smallcondensed]{svjour3}       % onecolumn "king-size"

\smartqed  % flush right qed marks, e.g. at end of proof

\usepackage{graphicx}
 \usepackage{mathptmx}      % use Times fonts if available on your TeX system
\usepackage{amssymb}
\usepackage{amsmath}
\usepackage{amsbsy}
\usepackage{bm}
\usepackage{color}
\usepackage{graphicx}
\usepackage[normalem]{ulem}
\usepackage{calrsfs}
\usepackage{cite}
\usepackage{hyperref}
\usepackage{subfigure}
\def\bal#1\eal{\begin{align}#1\end{align}}
\newcommand{\beq}{\begin{equation}}
\newcommand{\eeq}{\end{equation}}
\newcommand{\bq}{\begin{eqnarray}}
\newcommand{\eq}{\end{eqnarray}}
\newcommand{\bqn}{\begin{eqnarray*}}
\newcommand{\eqn}{\end{eqnarray*}}
\newcommand{\nn}{\nonumber\\}
\newcommand{\ex}{\mathrm{ex}}
\newcommand{\rmd}{\mathrm{d}}
\newcommand{\rme}{\mathrm{e}}
\newcommand{\rmi}{\mathrm{i}}

\newcommand{\GG}{\mathcal{G}}
\newcommand{\Xbeta}{X_\beta}

\begin{document}

\title{Triangle-Well and Ramp Interactions in One-Dimensional Fluids: A Fully Analytic Exact Solution}

\titlerunning{Triangle-Well and Ramp Interactions in One-Dimensional Fluids}        % if too long for running head

\author{Ana M. Montero         \and
        Andr\'es Santos
}


\institute{A. M. Montero \at
              Aachen Institute for Advanced Study in Computational Engineering Science, RWTH Aachen University, Schinkelstra{\ss}e 2, 52062 Aachen, Germany \\
                           %\email{}           %  \\
           \and
           A. Santos \at
              Departamento de F\'{\i}sica, Universidad de Extremadura,  06006 Badajoz, Spain \and Instituto de Computaci\'on Cient\'{\i}fica  Avanzada (ICCAEx), Universidad de Extremadura,  06006 Badajoz, Spain\\
                           \email{andres@unex.es}
}


\maketitle

\begin{abstract}
The exact statistical-mechanical solution for the equilibrium properties, both thermodynamic and structural, of one-dimensional fluids of particles interacting via the triangle-well and the ramp potentials is worked out. In contrast to previous studies, where the radial distribution function $g(r)$ was obtained numerically from the structure factor by Fourier inversion, we provide a fully analytic representation of $g(r)$ up to any desired distance. The solution is employed to perform an extensive study of the equation of state, the excess internal energy per particle, the residual multiparticle entropy, the structure factor, the radial distribution function, and the direct correlation function. In addition, scatter plots of the bridge function versus the indirect correlation function are used to gauge the reliability of the hypernetted-chain, Percus--Yevick, and Martynov--Sarkisov closures. Finally, the Fisher--Widom and Widom lines are obtained in the case of the triangle-well model.

 %\PACS{05.20.-y \and 05.20.Jj \and 05.70.-a \and 61.20.Ne \and 65.20.Jk}
\end{abstract}

\section{Introduction}
%%%%%%%%%%%%%%%%%%%%%%%%%%%%%%%%%%%%%%%%%%%%%%%%%%%%%%%%%%%%%%%%%%%%%%%%%%%%%%
\label{sec:introduction}

The number of statistical-mechanical models  lending themselves to exact solutions is very limited \cite{B08,M94}. One of the classes of equilibrium systems allowing for an exact treatment is made of particles  confined in one-dimensional  geometries with interactions restricted to first nearest
neighbors \cite{BNS09,BOP87,HGM34,HC04,KT68,K55b,K91,LPZ62,LZ71,R91,N40a,N40b,P76,P82,SZK53,S07,T42,T36}, as well as other one-dimensional systems, such as the one- and two-component plasma, the Kac--Backer model,  isolated self-gravitating systems,  interacting fermions and bosons, or the Toda lattice
\cite{F16,F17,M94,R71b}.
Most one-dimensional fluids with particles interacting beyond first nearest neighbors, however,  do not admit an exact solution and one needs to resort to approximations  \cite{F10b,FGMS10,FS17,SFG08}.
Even though relevant questions can be addressed  in a lattice gas or Ising model context \cite{BOP87,SHK18}, here we will focus on spatially continuous fluids.

The importance of exact statistical-mechanical solutions, even in conditions of one-dimensional confinement, cannot be overemphasized \cite{M94}. Not only do they represent academically  important examples of statistical-mechanical methods at work \cite{B83b,B89,BB83a,BB83d,HB08,LNP_book_note_13_08,S14,LNP_book_note_15_05_1,LNP_book_note_15_06_1,S16}, but they also provide insights into some of the expected general properties in unconfined geometries, or can be exploited as a benchmark for approximations \cite{ACE17,AE13,B84,BB83c,BB83b,BS86,BE02,S07,S07b} or simulation methods \cite{BB74,LK18}.
Moreover, since one-dimensional systems can be seen as three-dimensional systems confined in
a very narrow tube, they find a wide range of applications in physically important situations such as biological ion channels \cite{DNHEG08}, binding of proteins on capillary walls \cite{CPK15},  or carbon nanotubes \cite{Ketal11,MCH11}.

\begin{figure}%[htbp]
	\centering
	\subfigure[Triangle well]{\includegraphics[height=4cm]{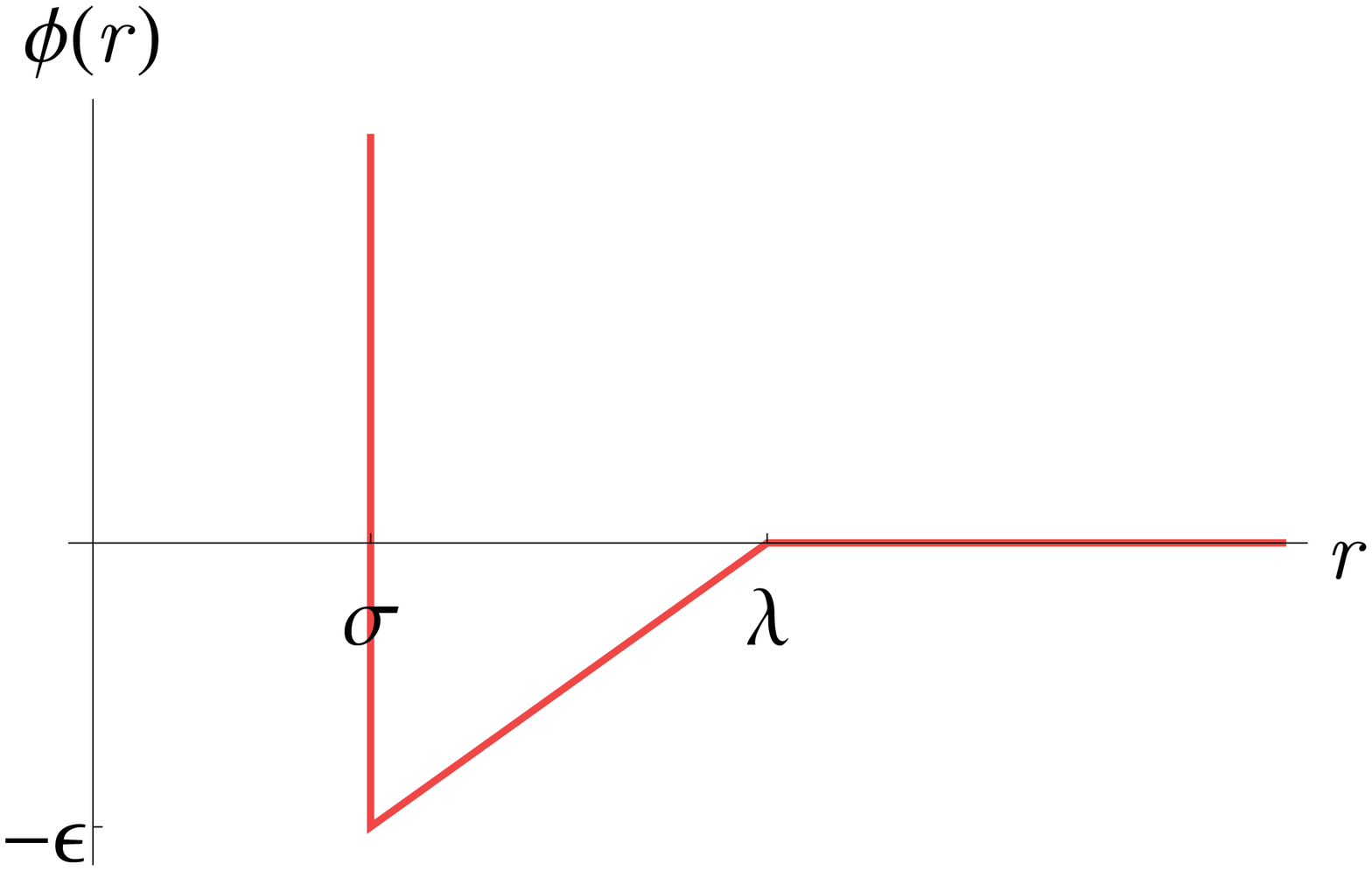} \label{trianglewell}}
	\subfigure[Ramp]{\includegraphics[height=4cm]{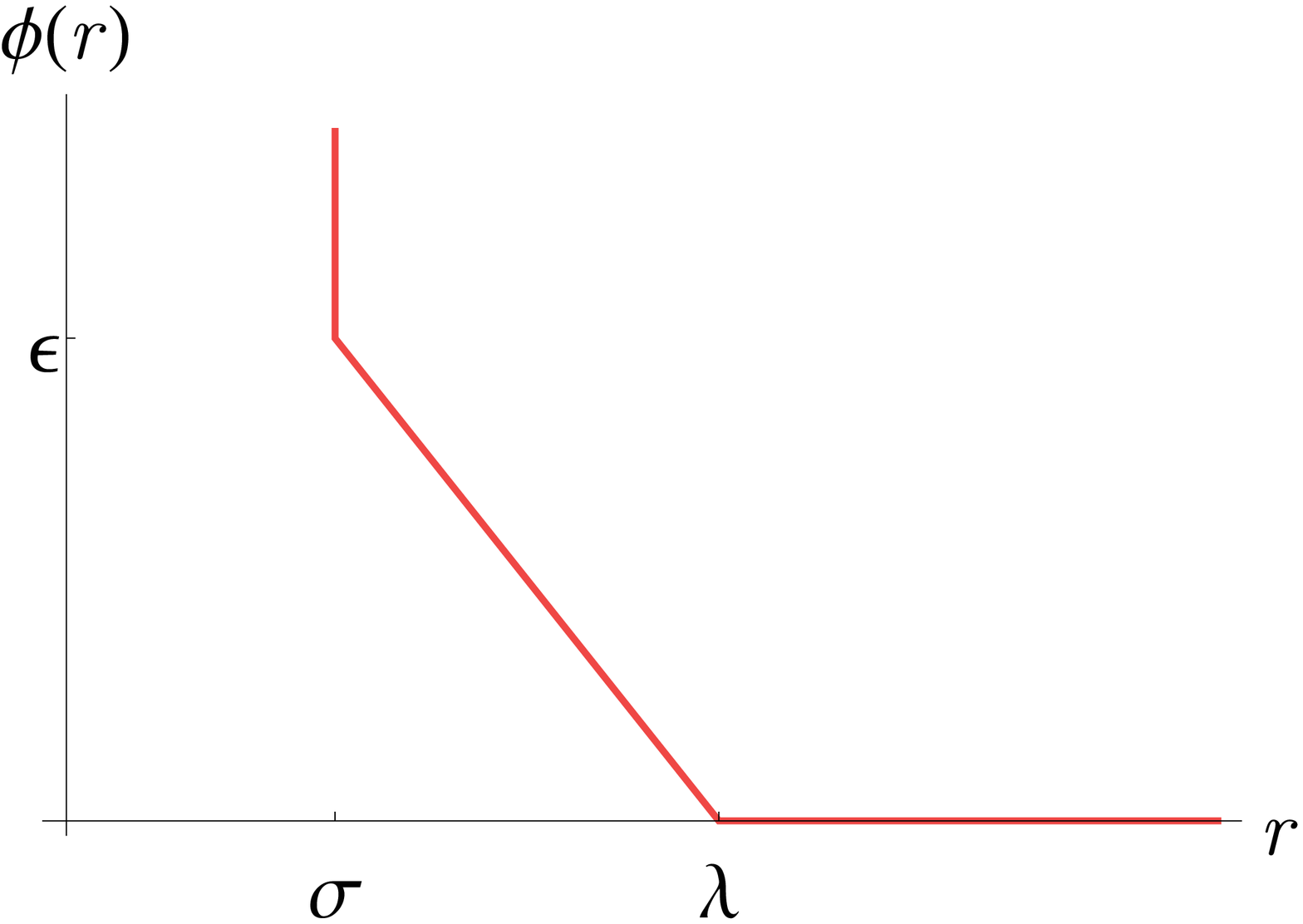} \label{ramp}}
	\caption{Sketch of (a) the triangle-well potential and (b) the ramp potential. }
	\label{studied_potentials}
\end{figure}

The aim of this paper is to derive  analytically the exact thermodynamic and structural properties of one-dimensional fluids of particles interacting via continuous potential functions of the form
\begin{equation}\label{eq:TW}
\phi (r) =
 \begin{cases}
 \infty & \text{if $r< \sigma$}, \\
 \displaystyle{-\epsilon\frac{ \lambda-r }{\lambda-\sigma}} & \text{if $\sigma < r \leq \lambda$}, \\
 0 & \text{if $r \ge \lambda$},
 \end{cases}
 \end{equation}
where $\sigma$ is the hard-core diameter of the particles and $\lambda$ is the range of the interactions. Apart from these two length scales, the potential includes an energy scale $\epsilon=-\phi(r=\sigma^+)$. If $\epsilon>0$, Eq.\ \eqref{eq:TW} defines the so-called triangle-well potential, as sketched in Fig.\ \ref{trianglewell}. This one-dimensional model was already studied by Nagamiya in 1940 \cite{N40a,N40b}. Its main importance lies in the fact that it represents the effective Asakura--Oosawa colloid-colloid depletion potential \cite{AO54,AO58,V76}
in a (one-dimensional) colloid-polymer
mixture in which the colloids are modeled by hard rods and the polymers are
treated as ideal particles excluded from the colloids by a certain distance \cite{BE02}.
In such a case, $\lambda=\sigma+\sigma_p$ and $\epsilon=k_BT \sigma_p z_p$, where $\sigma$ is the length of hard-rod colloids, $\sigma_p$ is the  length of ideal polymer coils, $k_B$ is the Boltzmann constant, $T$ is the temperature, and $z_p$ is the fugacity of ideal polymers.
This one-dimensional model has recently been used by Archer et al.\ \cite{ACE17} to assess the performance of the standard mean-field density functional theory. In their study, the authors obtained the pair correlation function by a numerical inversion of its exact Fourier transform, which gave rise to spurious oscillations in the neighborhood of $r=\sigma$. One of the goals of the present work is to express the pair correlation function in a fully analytic form, so those numerical instabilities will be absent.

By formally setting $\epsilon\to -\epsilon$ in Eq.\ \eqref{eq:TW}, the tail of the potential changes from attractive to repulsive and one has the purely repulsive ramp potential, as depicted in Fig.\ \ref{ramp}. This is an example of core-softened pair potentials, which generally present water-like anomalies \cite{LAMM07}. Although the triangle-well and the ramp potentials are physically very different, the exact solution to the former model immediately applies to the latter one by the formal change $\epsilon\to -\epsilon$.

The remainder of this paper is organized as follows. Section \ref{sec2} presents the main equations describing the exact properties of a one-dimensional fluid of particles which interact only with their nearest neighbors. The solution is explicitly worked out for the triangle-well and ramp potentials in Sec.\ \ref{sec3}, where the thermodynamic and structural properties, as well as the residual multiparticle entropy, are analyzed. Section \ref{sec4} is devoted to the assessment of three of the most popular approximate closures. The Fisher--Widom and Widom lines for the triangle-well fluid are obtained in Sec.\ \ref{sec5}. Finally, the main conclusions of the paper are summarized in Sec.\ \ref{sec6}.

\section{Exact Solution. General Framework}
\label{sec2}
Let us consider a one-dimensional system of $N$ particles in a box of
length $L$ (so that the number density is $n=N/L$) subject to a pair
interaction potential $\phi(r)$ such that (i) $\lim_{r\to 0}\phi(r)=\infty$, thus implying that the   {order} of the particles in the line does not change, and (ii) each particle interacts {only} with its first nearest neighbors.

The statistical-mechanical properties of the above model can be exactly obtained in the isothermal-isobaric ensemble \cite{SZK53,S16,S13,T42} where the thermodynamic state is characterized by temperature $T$ (or, equivalently $\beta\equiv 1/k_BT$) and pressure $p$. The key quantity in the solution is the Laplace transform of the Boltzmann factor $e^{-\beta\phi(r)}$, namely
\begin{subequations}
\label{2.1&2.2}
\beq
\label{2.1}
\Omega_\beta(s)\equiv\int_0^\infty \rmd r\, \rme^{-sr}\rme^{-\beta\phi(r)}.
\eeq
Its derivatives with respect to $s$ and $\beta$ are
\beq
\label{2.2}
\Omega_\beta'(s)\equiv \frac{\partial\Omega_\beta(s)}{\partial s}=-\int_0^\infty \rmd r\,  \rme^{-sr}r\rme^{-\beta\phi(r)},\quad
\Upsilon_\beta(s) \equiv - \frac{\partial \Omega_\beta(s)}{\partial \beta} = \int_0^{\infty} \rmd r\, \rme^{-sr}\phi(r) \rme^{-\beta\phi(r)}.
\eeq
\end{subequations}

It can be proved that the Gibbs free energy $G$, the  internal energy $U$, and the  Helmholtz free energy  $F$  are given by \cite{M17,S16}
 \begin{subequations}
 \label{eq:gibbs-omega}
 \begin{equation}
 \label{2.2a}
 	\frac{\beta G(N,p,T)}{N}=  \ln \frac{\Lambda_\beta}{\Omega_\beta(\beta p)} ,
 \quad
 	\frac{\beta U(N,p,T)}{N} =\frac{1}{2}+ \frac{\beta \Upsilon_\beta(\beta p)}{\Omega_\beta(\beta p)} ,
 \end{equation}
 \beq
 \frac{\beta F(N,p,T)}{N}=\ln \frac{\Lambda_\beta}{\Omega_\beta(\beta p)}
+\frac{\beta p\Omega'_\beta(\beta p)}{\Omega_\beta(\beta p)}.
\eeq
 \end{subequations}
Here, $\Lambda_\beta=h\sqrt{\beta/2\pi m}$ is the thermal de Broglie wavelength (where $h$ is the Planck constant and $m$ is the mass of a particle).
The equation of state, relating  number density $n$ with temperature $T$ and pressure $p$ is
\beq
    n(p,T)=-\frac{\Omega_\beta(\beta p)}{\Omega_\beta'(\beta p)}.
    \label{6.15}
    \eeq
The entropy $S$ is simply given as $S/k_B=\beta U-\beta F$. Thus, the \emph{excess} entropy per particle (relative to an ideal gas with the same temperature and density) is
\beq
\label{sex}
s_\ex\equiv\frac{S-S_{\text{ideal}}}{N k_B}=\beta u_\ex+Z-1+\ln\left[n\Omega_\beta(\beta p)\right],
\eeq
where $u_\ex\equiv U/N-k_BT/2$ is the excess internal energy per particle and $Z\equiv \beta p/n$ is the compressibility factor.

Equations \eqref{eq:gibbs-omega}--\eqref{sex} provide the main thermodynamic quantities. Additionally, the structural properties are described by the radial distribution function $g(r)$. Its Laplace transform is exactly given by \cite{S16}
   \beq
   \GG(s)\equiv \int_0^\infty \rmd r \, \rme^{-rs} g(r)=\frac{1}{n(p,T)}\frac{\Omega_\beta(s+\beta p)}{\Omega_\beta(\beta p)}\left[1-\frac{\Omega_\beta(s+\beta p)}{\Omega_\beta(\beta p)}\right]^{-1}.
   \label{6.6}
  \eeq
The Fourier transform of the total correlation function $h(r)\equiv g(r)-1$ can be directly obtained from $\GG(s)$ as
\beq
\widetilde{h}(k)\equiv \int_0^\infty\rmd r\, \left(\rme^{-\rmi kr}+\rme^{\rmi kr}\right)h(r)=\left[\GG(s)+\GG(-s)\right]_{s=\rmi k}.
\label{6.17a}
\eeq
{}From $\widetilde{h}(k)$ we can obtain the Fourier transforms of the direct and indirect correlation functions, as well as the static structure factor as
\beq
\label{2.7}
\widetilde{c}(k)=\frac{\widetilde{h}(k)}{1+n\widetilde{h}(k)},\quad \widetilde{\gamma}(k)\equiv\widetilde{h}(k)-\widetilde{c}(k)=\frac{n\widetilde{h}^2(k)}{1+n\widetilde{h}(k)},\quad\widetilde{S}(k)=1+n\widetilde{h}(k).
\eeq

For later use, we introduce here the residual multiparticle entropy (RMPE) $\Delta s$ as \cite{G08,GG92,SSG18}
\beq
\label{2.9a}
\Delta s\equiv s_\ex-s_2, \quad s_2=-n\int_0^\infty \rmd r\, \left[g(r)\ln g(r)-h(r)\right],
\eeq
where $s_2$
is the pair correlation contribution to the excess entropy per particle (in units of $k_B$).
Thus, $\Delta s$ represents an integrated measure of the importance of more-than-two-particle density
correlations in the overall entropy balance.
\section{Exact Solution. Triangle-well and Ramp Potentials}
\label{sec3}
In the special case of the potential \eqref{eq:TW}, the Laplace functions \eqref{2.1&2.2} become
\begin{subequations}
\label{eq:Omega-Upsilon}
 \begin{equation}
 \label{eq:Omega}
\Omega_\beta(s)= \frac{\Xbeta \rme^{-s}}{a_\beta+s} \left[1+\frac{a_\beta \rme^{-(\lambda-1)s}}{\Xbeta s} \right] ,
\end{equation}
\begin{equation}
\label{eq:Omega'}
 	\Omega_\beta'(s)= -\frac{\lambda  \rme^{-\lambda s}}{s}-\frac{\rme^{-\lambda s}}{s^2}+\frac{\lambda  \rme^{-\lambda s}-\Xbeta\rme^{-s}}{a_\beta+s}+\frac{\rme^{-\lambda  s}-\Xbeta\rme^{-s}}{\left(a_\beta+s\right)^2} ,
 \end{equation}
\begin{equation}
\label{eq:Upsilon}
	\Upsilon_\beta(s) = -\frac{\epsilon}{a_\beta+s} \left[\Xbeta\rme^{-s} + \frac{\rme^{-\lambda s}-\Xbeta\rme^{-s}}{(\lambda-1)(a_\beta+s)}\right],
 \end{equation}
 \end{subequations}
 where henceforth we choose $\sigma=1$ as the length unit (although we will use $n^*\equiv n\sigma$ for the reduced density) and
 \begin{equation}
 \label{eq:Xa}
 	 \Xbeta \equiv \rme^{\beta \epsilon}, \quad  	a_\beta \equiv \frac{\beta \epsilon}{\lambda - 1}.
 \end{equation}

In the high-temperature limit ($\beta\epsilon\to 0$), $a_\beta\to 0$ and thus $\Omega_\beta(s)\to \rme^{-s}/s$, which is the hard-rod result. Interestingly, in the case of the ramp model ($\epsilon\to -\epsilon$), the low-temperature limit ($\beta\epsilon\to \infty$) implies $\Xbeta\to 0$ and $a_\beta\to-\infty$, so that $\Omega_\beta(s)\to \rme^{-\lambda s}/s$, i.e., the result corresponding to hard rods of length $\lambda$.
Finally, in the case of the triangle-well model, the combined limit $\beta\epsilon\to\infty$, $\lambda-1\to 0$ with fixed $\tau^{-1}\equiv \Xbeta/a_\beta$, leads to $\Omega_\beta(s)\to \rme^{-s}(\tau^{-1}+s^{-1})$, which corresponds to sticky hard rods with a stickiness parameter $\tau^{-1}$ \cite{B68,S16,YS93a}.

\subsection{Thermodynamic Quantities}
According to Eqs.\ \eqref{2.2a} and \eqref{6.15}, the excess internal energy per particle and the equation of state are
\begin{subequations}
\label{eq:energy&density}
\begin{equation}
\label{eq:energy-tw}
u_\ex(p,T)= -\epsilon \frac{1+\Xbeta \rme^{\beta p (\lambda-1) } \left[(a_\beta+\beta p) (\lambda -1)-1\right]}{
(a_\beta+\beta p) (\lambda -1)\left[a_\beta/\beta p + \Xbeta \rme^{\beta p (\lambda-1) } \right]},
\end{equation}
 \begin{equation}
\label{eq:density-tw}
	\frac{1}{n^*(p,T)} = 1+\frac{1}{\beta p+a_\beta}+\frac{\lambda -1+1/\beta p}{1+\beta p (\Xbeta/a_\beta)\rme^{\beta p(\lambda-1)} } .
\end{equation}
\end{subequations}

Since the exact analytic solution is expressed in the isothermal-isobaric ensemble, Eq.\ \eqref{eq:density-tw} gives $n$ as an explicit function of $\beta$ and $\beta p$. On the other hand, it is usually more convenient to choose $n$ and $T$ as the two independent thermodynamic variables. In that case, Eq.\ \eqref{eq:density-tw} can be interpreted as the transcendental equation giving $\beta p$ as a function of $(n,T)$. Its solution is easily obtained by using the hard-rod   expression, i.e., $\lim_{\beta\epsilon\to 0}\beta p=n/(1-n^*)$, as an initial estimate.
Moreover, the coefficients in the virial series
\beq
\label{virial_series}
\beta p=n+B_2(T)n^2+B_3(T)n^3+B_4(T)n^4+\cdots
\eeq
can be derived in a recursive way by inserting \eqref{virial_series} into the right-hand side of Eq.\ \eqref{eq:density-tw}, expanding in powers of $n$ and making all the coefficients beyond first order equal to zero. The first few coefficients are
\begin{subequations}
  \beq
  B_2=\lambda-\frac{\Xbeta-1}{a_\beta}, \quad B_3=\lambda^2-2\frac{(2\lambda-1)\Xbeta-\lambda}{a_\beta}+2\Xbeta\frac{\Xbeta-1}{a_\beta^2},
  \eeq
  \beq
  B_4=\lambda^3-3\frac{(7\lambda^2-6\lambda+1)\Xbeta-2\lambda^2}{2a_\beta}+3\Xbeta\frac{(5\lambda-3)\Xbeta-2(2\lambda-1))}{a_\beta^2}
  -\frac{\Xbeta^2(5\Xbeta-6)+1}{a_\beta^3}.
  \eeq
\end{subequations}
In the opposite limit of densities near close-packing ($n^*\to 1$), the pressure tends to infinity and the excess internal energy per particle tends to $-\epsilon$ as
\beq
\label{2.13}
\beta p\approx \frac{n}{1-n^*}-a_\beta,\quad u_\ex\approx -\epsilon\frac{\lambda n^*-1}{(\lambda-1)n^*}.
\eeq
This is obtained from Eqs.\ \eqref{eq:energy&density} by taking the limit $e^{\beta p(\lambda-1)}\to\infty$.

\begin{figure}%[htbp]
	\centering
	\subfigure[Triangle well]{\includegraphics[height=5cm]{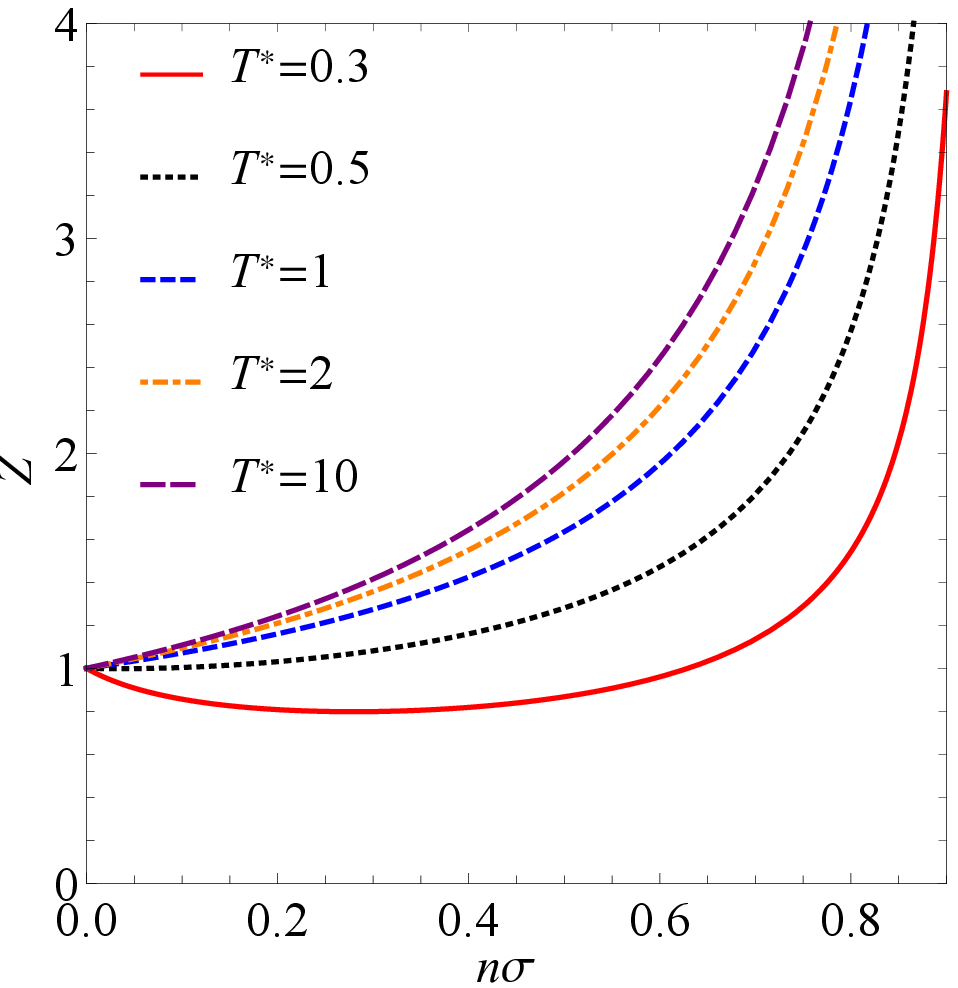} \label{well_Z}}\hspace{1cm}
	\subfigure[Ramp]{\includegraphics[height=5cm]{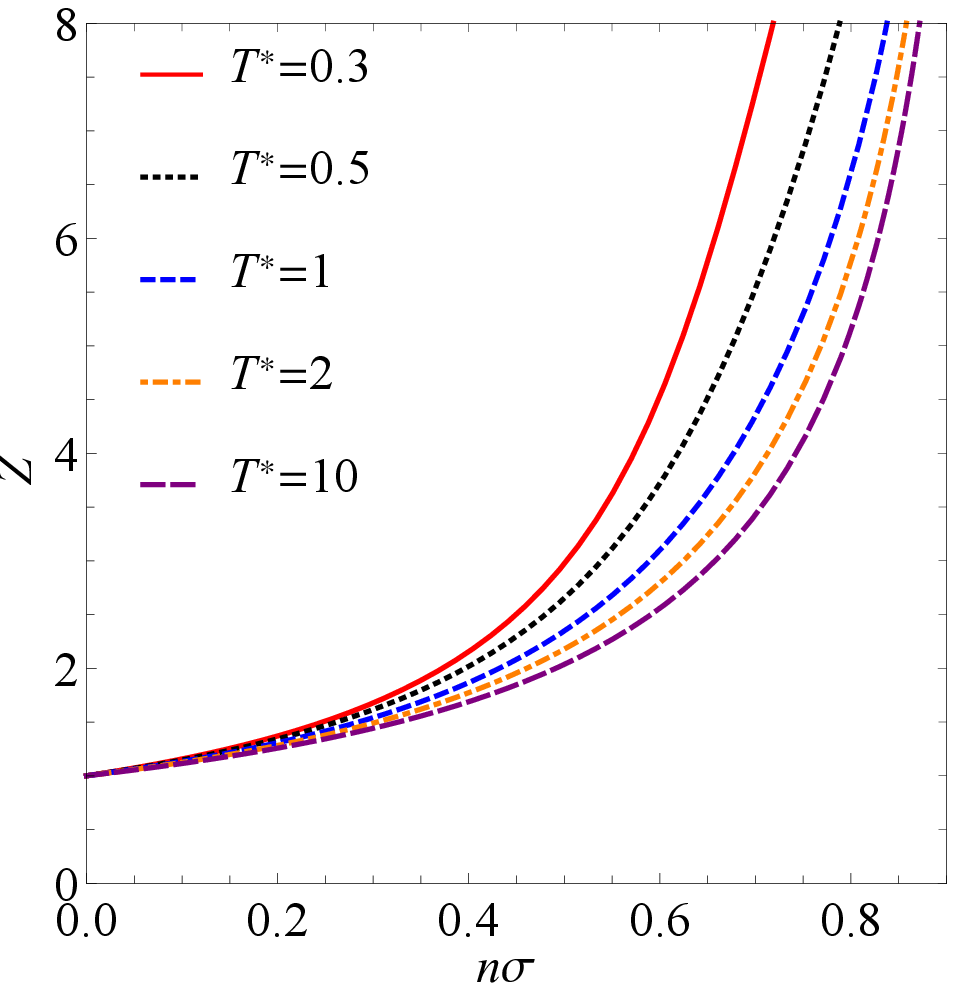} \label{ramp_Z}}
	\caption{Compressibility factor $Z$ versus density at several representative temperatures for (a) the triangle-well potential and (b) the ramp potential, in both cases with $\lambda=1.5$. }
	\label{fig:Z}
\end{figure}

\begin{figure}%[htbp]
	\centering
	\subfigure[Triangle well]{\includegraphics[height=5cm]{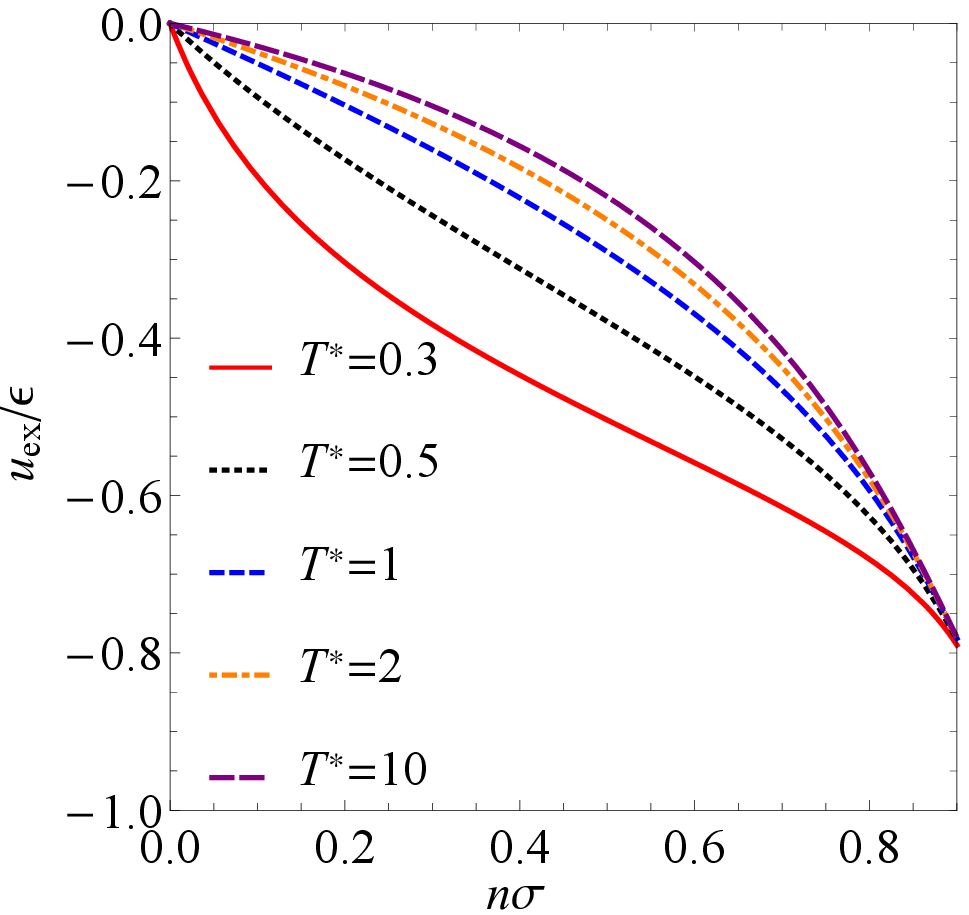} \label{well_uex}}\hspace{1cm}
	\subfigure[Ramp]{\includegraphics[height=5cm]{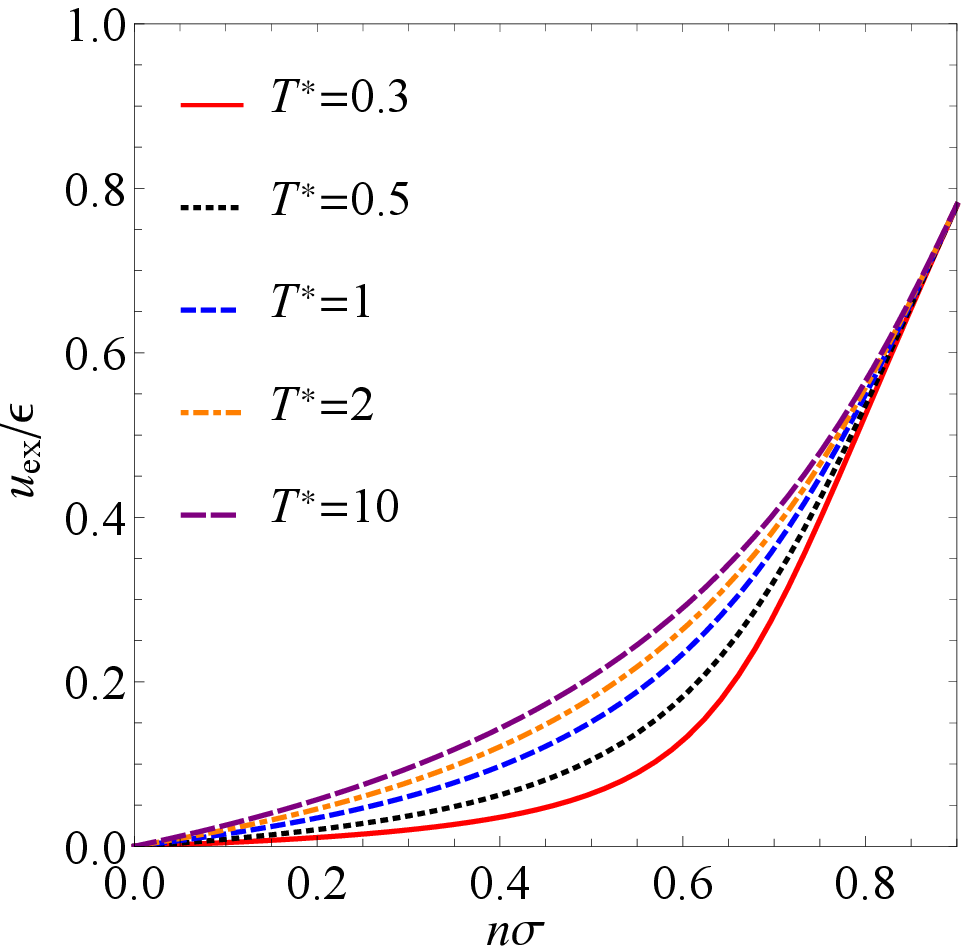} \label{ramp_uex}}
	\caption{Excess internal energy per particle $u_\ex$ versus density at several representative temperatures for (a) the triangle-well potential and (b) the ramp potential, in both cases with $\lambda=1.5$. }
	\label{fig:uex}
\end{figure}

Figure \ref{fig:Z} shows the compressibility factor $Z\equiv \beta p/n$ as a function of density at several representative values of the reduced temperature $T^*\equiv k_BT/\epsilon$. Panels (a) and (b) correspond to the triangle-well and ramp potentials, respectively, in both cases with $\lambda=1.5$. Note that in the case of the ramp potential the formal change $\epsilon\to-\epsilon$ needs to be made [see Fig.\ \ref{ramp}]. Thus, while $\Xbeta>1$ and $a_\beta>0$ in the triangle-well model, $\Xbeta<1$ and $a_\beta<0$ in the ramp model.
In the former model, if the temperature is low enough, so that $\Xbeta-1>a_\beta\lambda$, one has  $B_2(T)<0$ and thus the compressibility factor first decreases, then reaches a minimum, and finally  increases with increasing density; otherwise, $Z$ increases monotonically. On the other hand, in the purely repulsive ramp model $Z$ is always an increasing function. Note that the hard-rod compressibility factor $1/(1-n^*)$ is an upper (lower) bound in the triangle-well (ramp) model.

The excess internal energy per particle (in units of $\epsilon$) is shown in Fig.\ \ref{fig:uex}. We can observe that for densities $n\gtrsim 0.9$ the excess internal energy becomes practically independent of temperature, in agreement with Eq.\ \eqref{2.13}. Actually, $u_\ex/\epsilon=-(\lambda n^*-1)/(\lambda-1)n^*$ is an upper (lower) bound in the triangle-well (ramp) model.

\subsection{Structural Properties}

\begin{figure}%[htbp]
	\centering
	\subfigure[$n^*=0.6$, Triangle well]{\includegraphics[height=5cm]{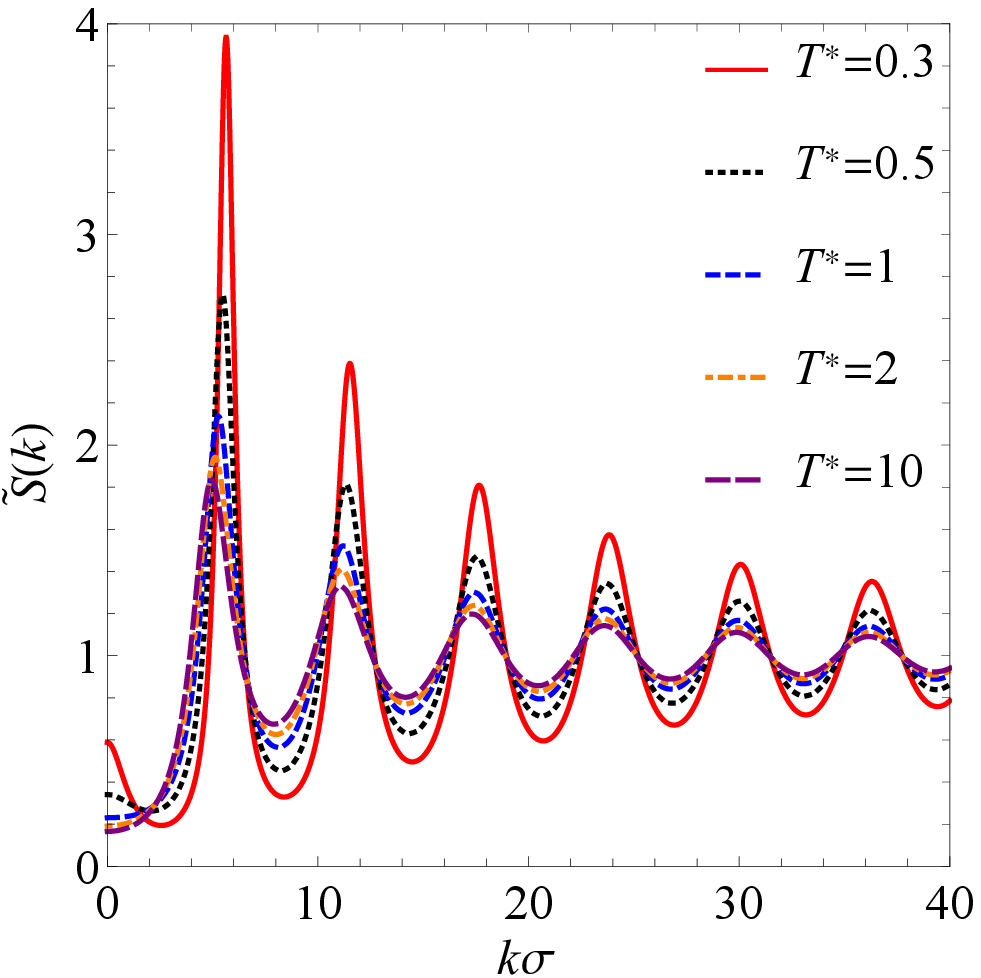} \label{Sk_T_well}}\hspace{1cm}
	\subfigure[$n^*=0.6$, Ramp]{\includegraphics[height=5cm]{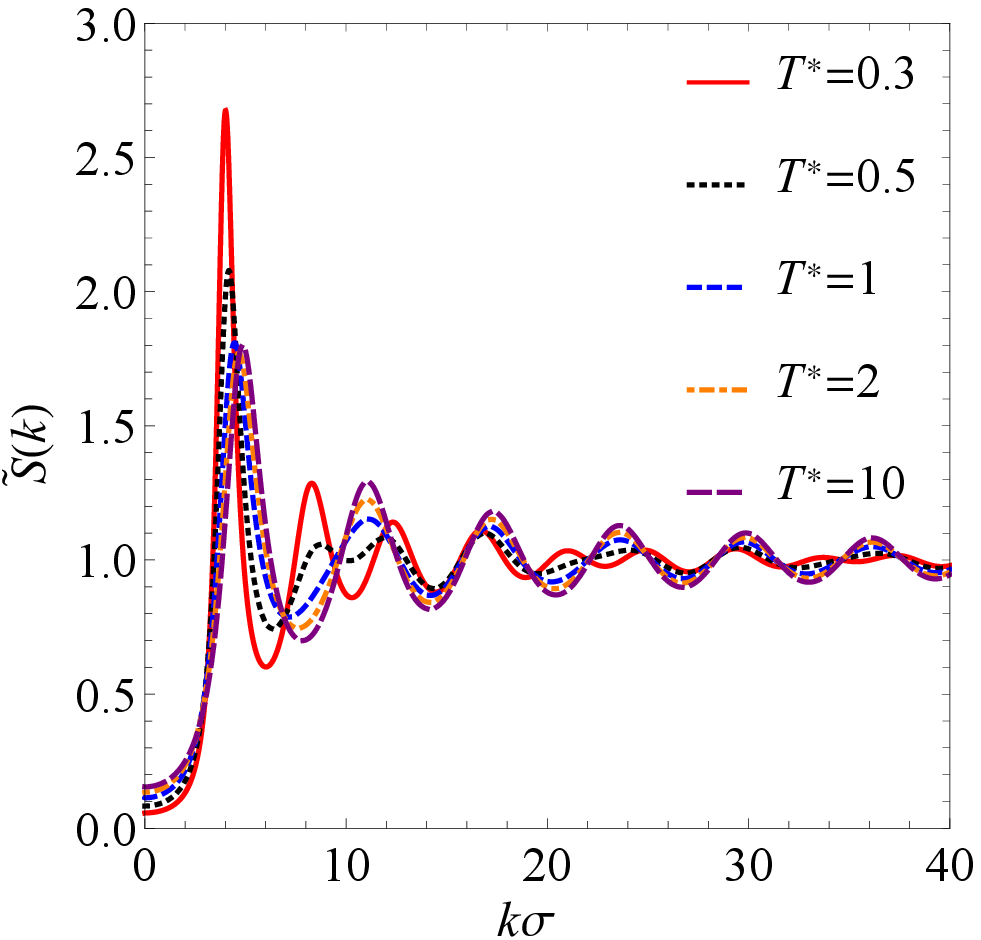} \label{Sk_T_ramp}}\\
\subfigure[$T^*=1$, Triangle well]{\includegraphics[height=5cm]{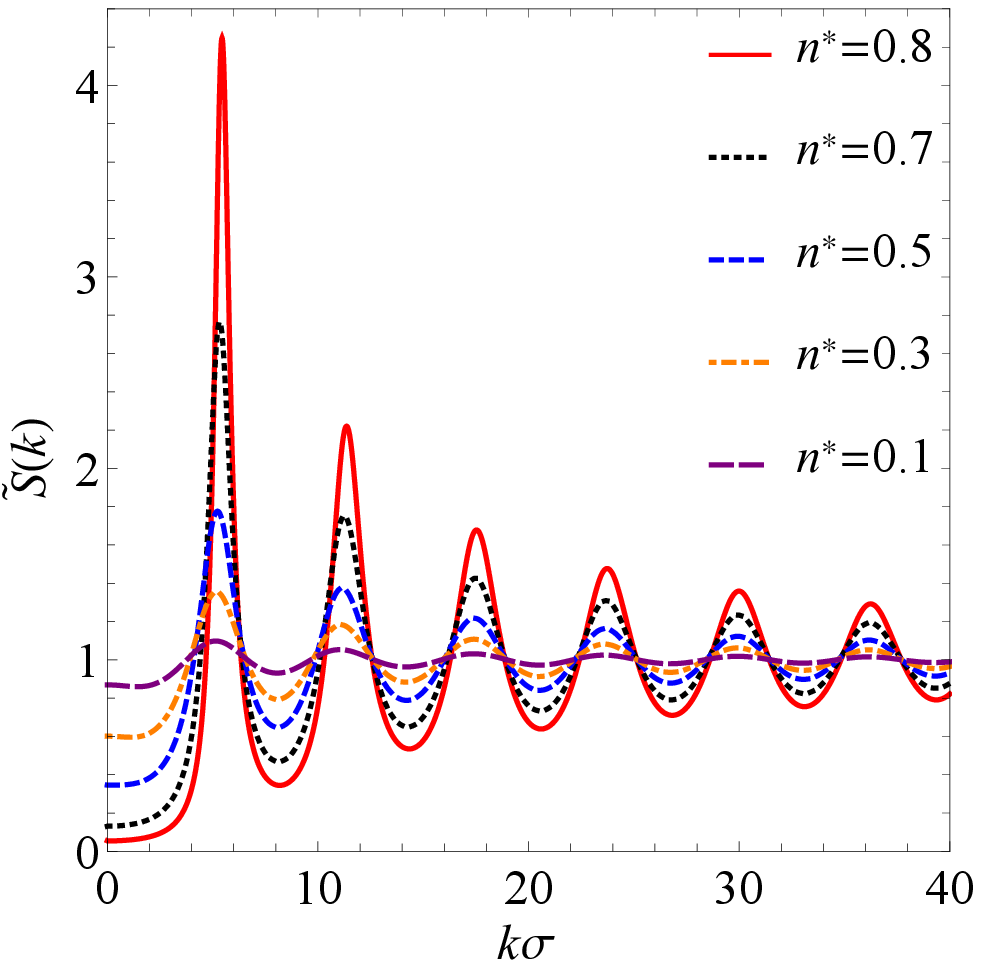} \label{Sk_n_well}}\hspace{1cm}
	\subfigure[$T^*=1$, Ramp]{\includegraphics[height=5cm]{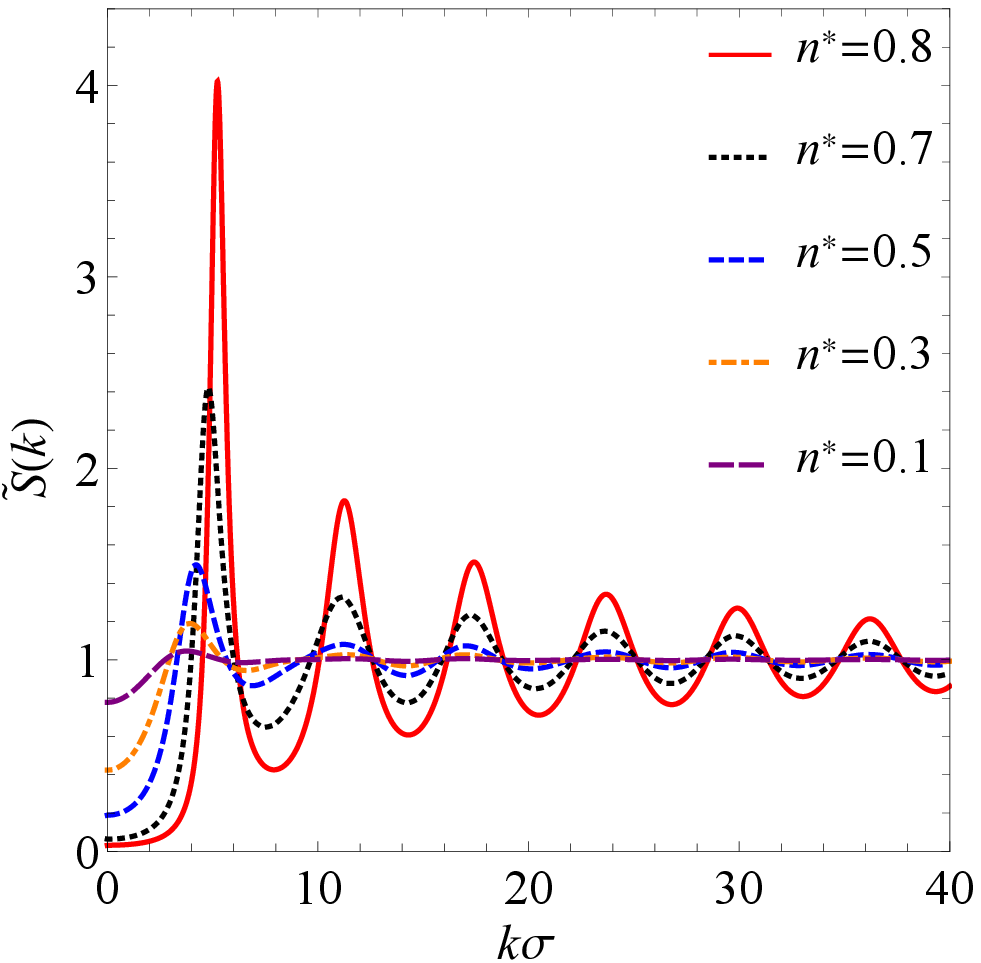} \label{Sk_n_ramp}}
	\caption{Static structure factor $\widetilde{S}(k)$ at several representative temperatures for a reduced density $n^*=0.6$ [panels (a) and (b)] and at several representative densities for a reduced temperature $T^*=1$ [panels (c) and (d)]. Panels (a) and (c) correspond to the triangle-well potential, while panels (b) and (d) correspond to the ramp potential, in all cases with $\lambda=1.5$. }
	\label{fig:Sk}
\end{figure}

\begin{figure}%[htbp]
	\centering
	\subfigure[$n^*=0.6$, Triangle well]{\includegraphics[height=5cm]{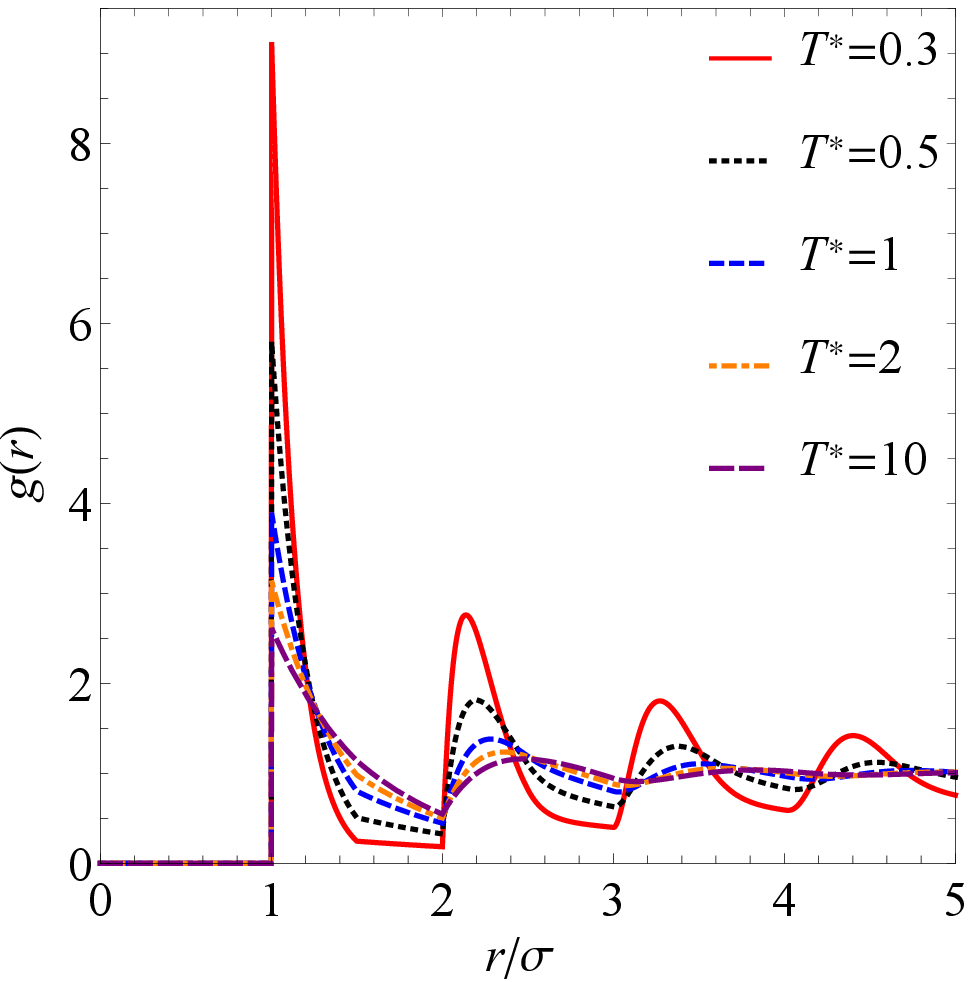} \label{gr_T_well}}\hspace{1cm}
	\subfigure[$n^*=0.6$, Ramp]{\includegraphics[height=5cm]{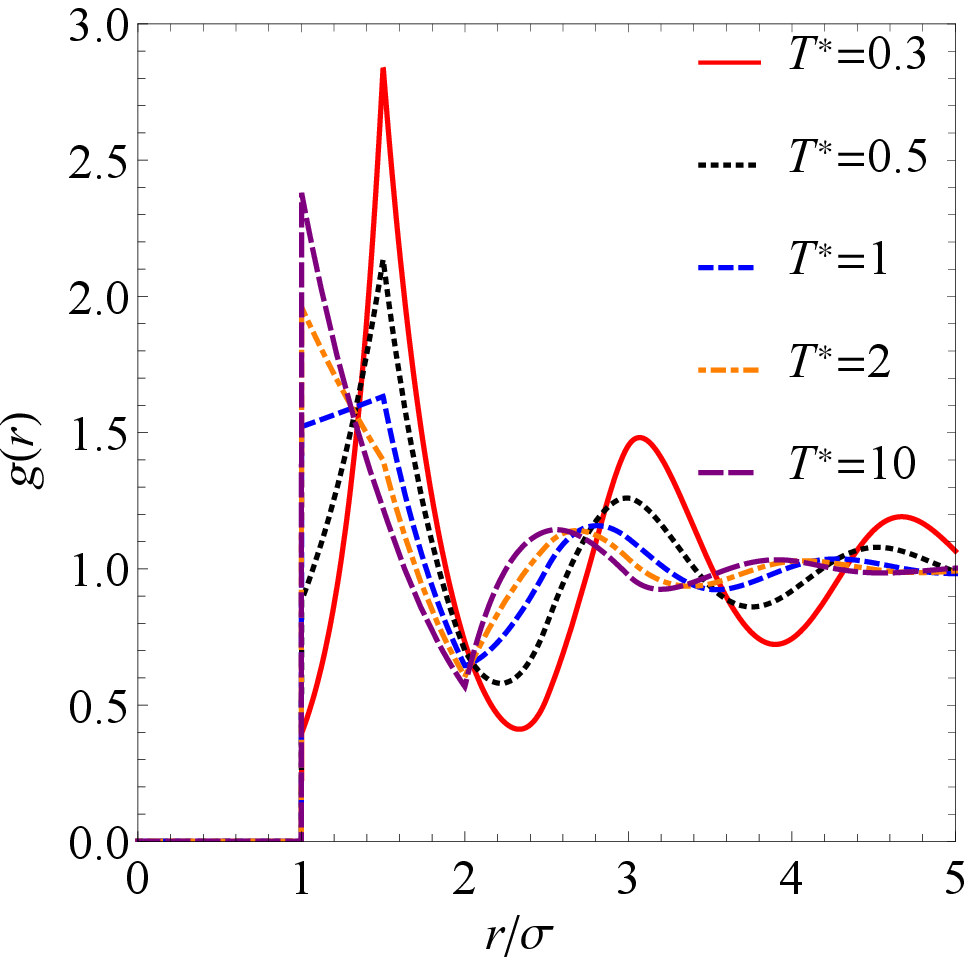} \label{gr_T_ramp}}\\
\subfigure[$T^*=1$, Triangle well]{\includegraphics[height=5cm]{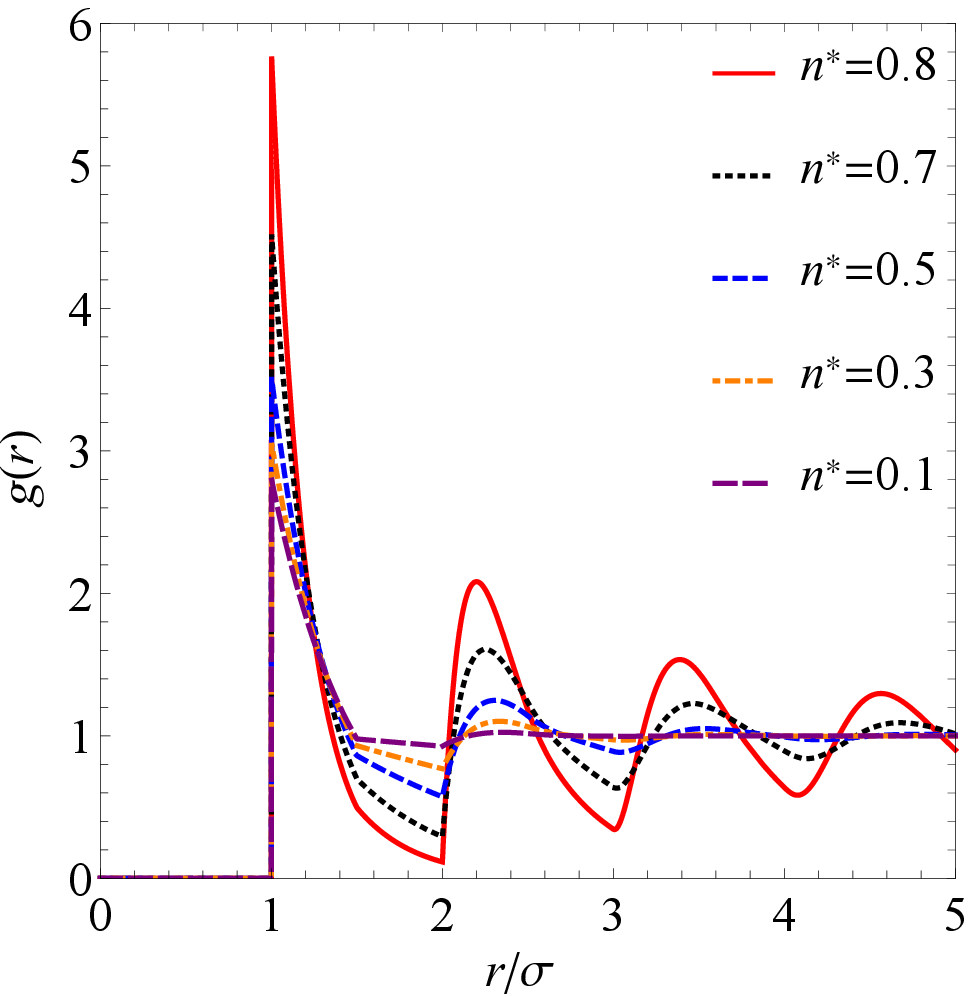} \label{gr_n_well}}\hspace{1cm}
	\subfigure[$T^*=1$, Ramp]{\includegraphics[height=5cm]{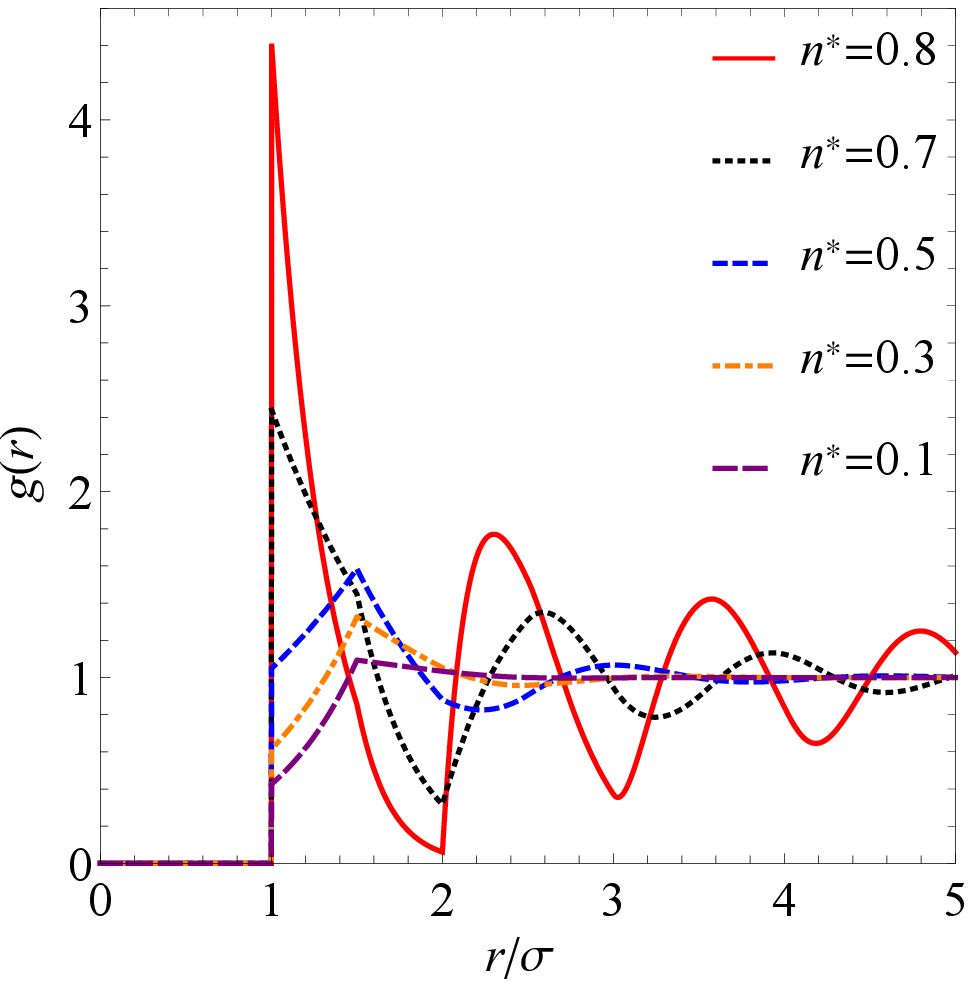} \label{gr_n_ramp}}
	\caption{Radial distribution function $g(r)$ at several representative temperatures for a reduced density $n^*=0.6$ [panels (a) and (b)] and at several representative densities for a reduced temperature $T^*=1$ [panels (c) and (d)]. Panels (a) and (c) correspond to the triangle-well potential, while panels (b) and (d) correspond to the ramp potential, in all cases with $\lambda=1.5$. }
	\label{fig:gr}
\end{figure}

\begin{figure}%[htbp]
	\centering
	\subfigure[$n^*=0.6$, Triangle well]{\includegraphics[height=5cm]{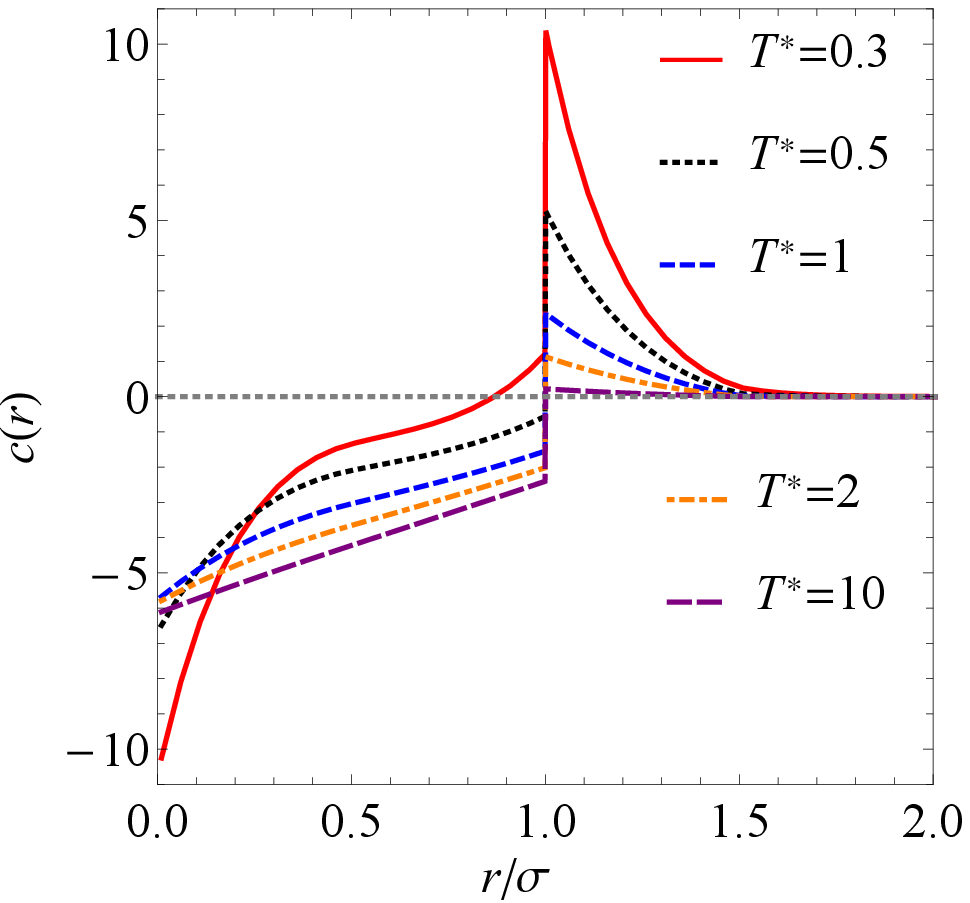} \label{cr_T_well}}\hspace{1cm}
	\subfigure[$n^*=0.6$, Ramp]{\includegraphics[height=5cm]{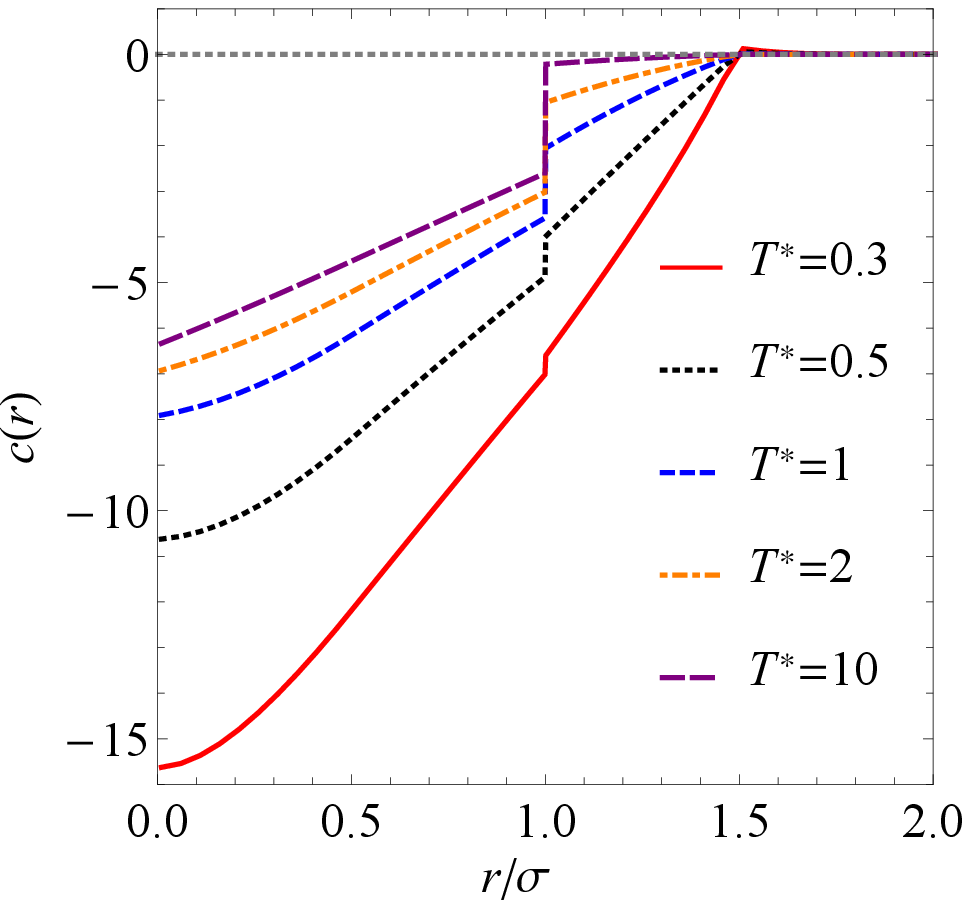} \label{cr_T_ramp}}\\
\subfigure[$T^*=1$, Triangle well]{\includegraphics[height=5cm]{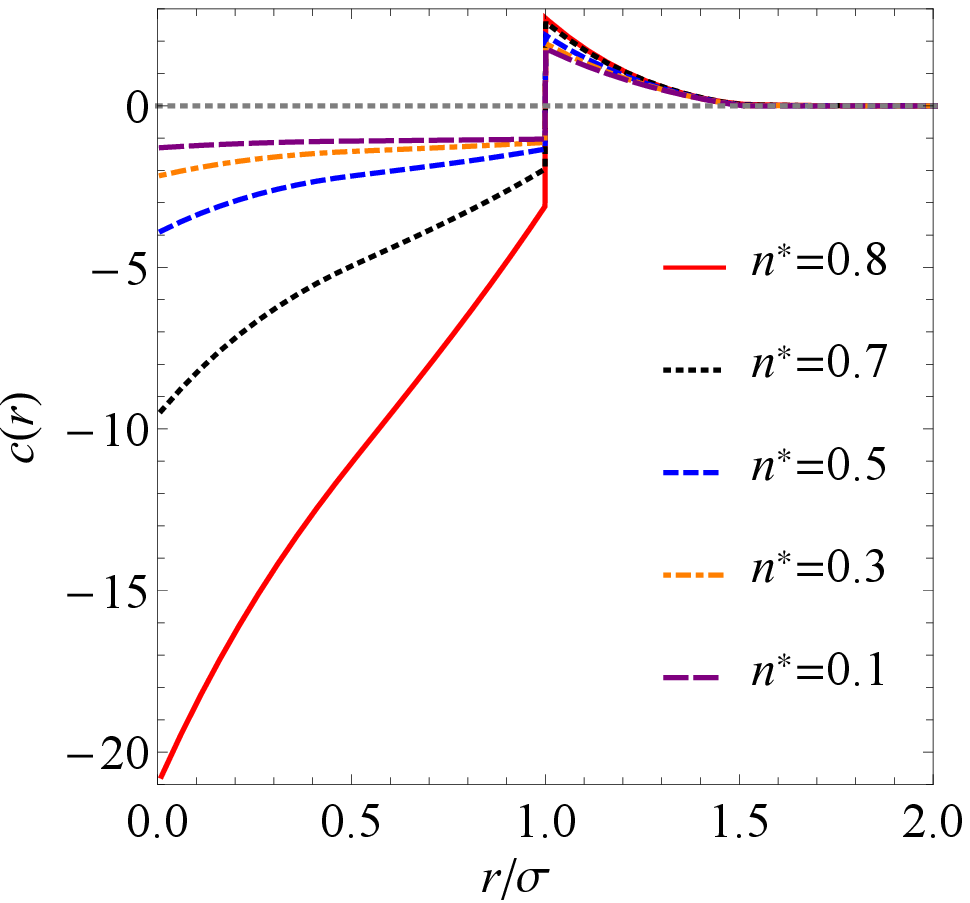} \label{cr_n_well}}\hspace{1cm}
	\subfigure[$T^*=1$, Ramp]{\includegraphics[height=5cm]{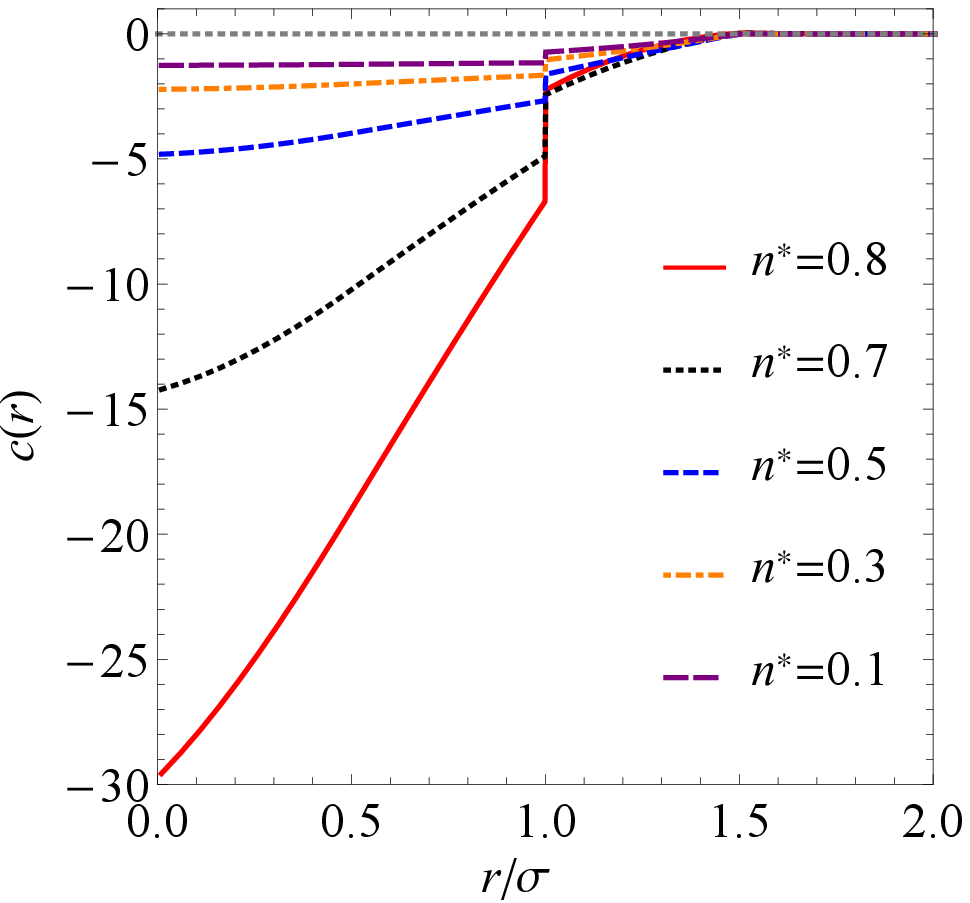} \label{cr_n_ramp}}
	\caption{Direct correlation function $c(r)$ at several representative temperatures for a reduced density $n^*=0.6$ [panels (a) and (b)] and at several representative densities for a reduced temperature $T^*=1$ [panels (c) and (d)]. Panels (a) and (c) correspond to the triangle-well potential, while panels (b) and (d) correspond to the ramp potential, in all cases with $\lambda=1.5$. }
	\label{fig:cr}
\end{figure}

The Laplace transform $\GG(s)$ is obtained by inserting Eq.\ \eqref{eq:Omega} into Eq.\ \eqref{6.6}. Next, the static structure factor $\widetilde{S}(k)$ is obtained in analytic form from Eqs.\ \eqref{6.17a} and \eqref{2.7}. Figure \ref{fig:Sk} shows $\widetilde{S}(k)$ for a few representative states. We can observe that the triangle-well fluid tends to have a more structured state than the purely repulsive ramp fluid. Moreover, while in the former case the locations of the maxima and minima are hardly dependent on density and/or temperature, this is not so in the ramp case, especially in what concerns the influence of temperature.

Now we turn our attention to the radial distribution function $g(r)$, which measures the structure of the fluid in real space. In principle, one can invert numerically either the analytic Laplace transform $\GG(s)$ \cite{AW92} or the analytic Fourier transform $\widetilde{h}(k)$ to obtain $g(r)$. However, either of those possibilities leads to numerical instabilities, especially around $r=\sigma$ \cite{ACE17}. In order to avoid those problems, we derive here an analytic expression for $g(r)$. The starting point is the formal series expansion of the last equality of Eq.\ \eqref{6.6} in powers of $\Omega_\beta(s+\beta p)$ \cite{M17,MS17}, i.e.,
\begin{equation}
	\GG(s)=
	\frac{1}{n} \sum_{\ell=1}^{\infty}      \left[ \frac{\Omega_\beta(s+\beta p)}{\Omega_\beta(\beta p)} \right]^\ell .
\end{equation}
Thus,
\beq
\label{g(r)}
g(r)=\frac{1}{n} \sum_{\ell=1}^{\infty}      \frac{\mathcal{L}^{-1}\left\{\left[\Omega_\beta(s+\beta p) \right]^\ell\right\}}{\left[\Omega_\beta(\beta p)\right]^\ell}  ,
\eeq
where $\mathcal{L}^{-1}\left\{\cdots\right\}$ denotes the inverse Laplace transform.
From the explicit expression \eqref{eq:Omega}, we have
\beq
\label{2.16}
	\left[\Omega_\beta(s + \beta p)\right]^\ell = \sum_{\ell_1=0}^{\ell} C_{\ell \ell_1} \rme^{-(\ell-\ell_1+\ell_1\lambda)s} F_{\ell \ell_1}(s) ,
\eeq
where we have called
\begin{equation}
\label{Fs}
	C_{\ell \ell_1} \equiv \binom{\ell}{\ell_1}a_\beta^{\ell_1}\Xbeta^{\ell-\ell_1} \rme^{-\beta p \ell}\rme^{-\beta p  (\lambda -1)\ell_1} ,\quad F_{\ell \ell_1}(s) \equiv \frac{1}{ (a_\beta+\beta p+s)^\ell(\beta p+s)^{\ell_1}} .
\end{equation}

Combination of Eqs.\ \eqref{g(r)} and \eqref{2.16} yields
\beq
\label{g(r)2}
g(r)=\frac{1}{n^*} \sum_{\ell=1}^{[r]}  \frac{1}{\left[\Omega_\beta(\beta p)\right]^\ell} \sum_{\ell_1=0}^{\ell} C_{\ell \ell_1}  f_{\ell \ell_1}(r-\ell+\ell_1-\ell_1\lambda) ,
\eeq
where $f_{\ell\ell_1}(r)\equiv\mathcal{L}^{-1}\left\{F_{\ell\ell_1}(s)\right\}$. This function  is derived in Appendix \ref{appA} [see Eq.\ \eqref{A16}].
In Eq.\ \eqref{g(r)2}, the upper limit $\ell\to\infty$ in the first summation has been replaced by $\ell=[r]$, where $[r]$ denotes the integer part of $r$, due to the Heaviside step function appearing in Eq.\ \eqref{A16}. Equation \eqref{g(r)2} provides an exact explicit expression of $g(r)$ without any numerical Laplace or Fourier inversion. In particular, in the interval $1<r<\lambda$,
\beq
\label{gclose}
g(r)=\frac{\Xbeta}{n^*\Omega_\beta(\beta p)}\rme^{-(a_\beta+\beta p)(r-1)}=\frac{\Xbeta}{n^*\Omega_\beta(\beta p)}\rme^{-\beta\phi(r)}\rme^{-\beta p(r-1)},\quad 1<r<\lambda.
\eeq

Figure \ref{fig:gr} shows $g(r)$ for the same cases as in Fig.\ \ref{fig:Sk}. As in the case of the structure factor, we observe that the location of the maxima and minima is much more sensitive to the values of temperature and density in the ramp model than in the triangle-well model. Additionally, an interesting transition from a negative slope $g'(1^+)$ to a positive slope as temperature and/or density decrease is present in ramp fluids. This can be easily understood from Eq.\ \eqref{gclose}. Since in the ramp potential (where $\epsilon\to -\epsilon$) $a_\beta<0$, one has $g'(1^+)>$ when $\beta p<|a_\beta|$. From the equation of state \eqref{eq:density-tw}, it is possible to see that the above condition is equivalent to
\beq
n^*\leq \left[1+(\lambda-1)\frac{T^{*2}+T^*+2}{2(T^*+1)}\right]^{-1}.
\eeq

To complement the study of the structural properties, we now turn our attention to the direct correlation function $c(r)$. In contrast to the case of $g(r)$, no analytic expression for $c(r)$ seems to be possible. Since $\widetilde{c}(k)$ is explicitly known [see Eqs.\ \eqref{6.17a} and \eqref{2.7}], one could obtain $c(r)$ by numerical Fourier inversion. On the other hand, given that $c(r)$ is discontinuous at $r=1$, while $\gamma(r)$ is continuous, it is more convenient to obtain the indirect correlation function from the numerical Fourier inversion $\gamma(r)=\pi^{-1}\int_0^\infty \rmd k\,\cos(kr)\widetilde{\gamma}(k)$ and then use the relation $c(r)=h(r)-\gamma(r)$. The results are shown in Fig.\ \ref{fig:cr}. We observe that, while $c(r)$ is generally negative in the region $r<1$ for both classes of fluids, in the region $r>1$ one typically has $c(r)>0$ and $c(r)<0$ for the triangle-well and ramp potentials, respectively. It is worthwhile noticing that $c(r)$ becomes  negligible  beyond the range of the potential ($r>\lambda$), even at low temperatures and/or high densities.

\begin{figure}%[htbp]
	\centering
	\subfigure[Triangle well]{\includegraphics[height=5cm]{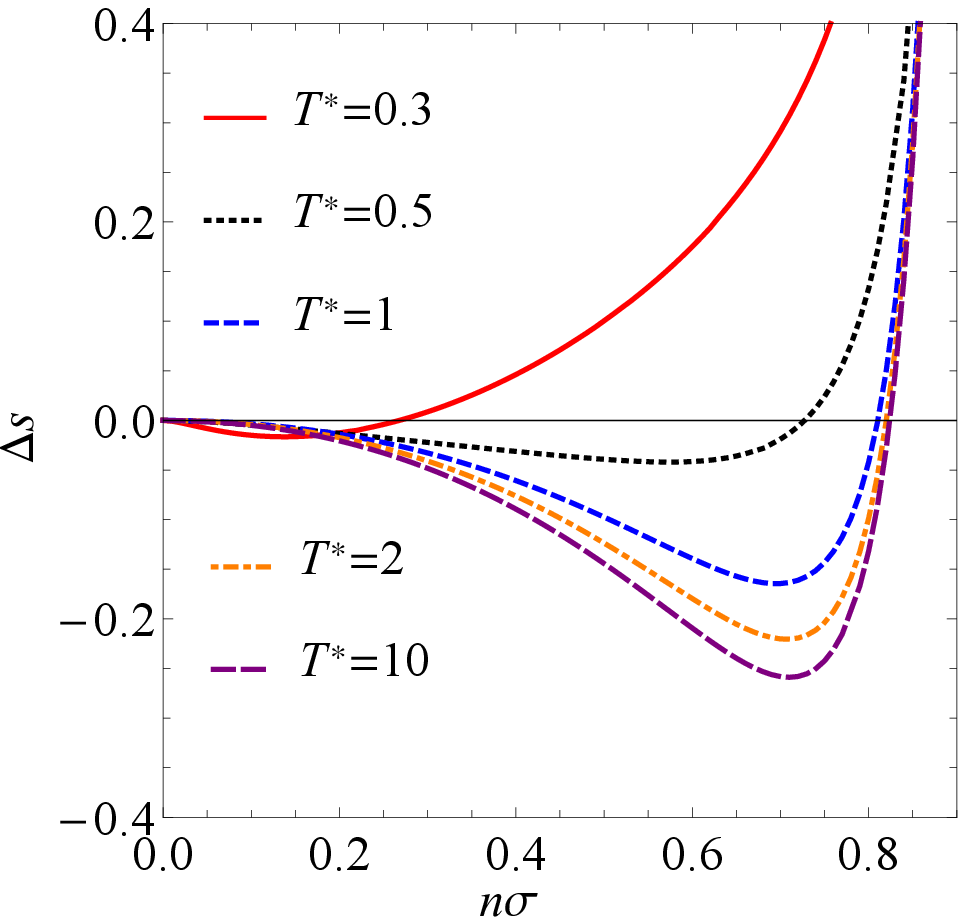} \label{well_RMPE}}\hspace{1cm}
	\subfigure[Ramp]{\includegraphics[height=5cm]{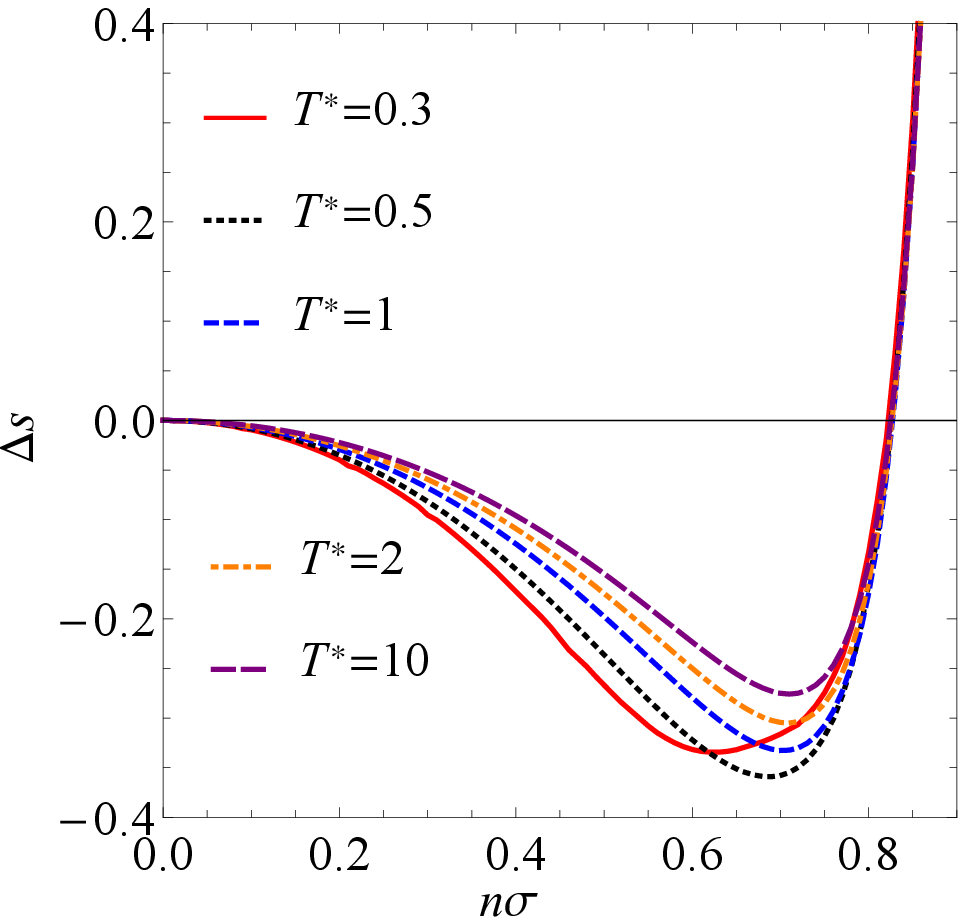} \label{ramp_RMPE}}
	\caption{Residual multiparticle entropy $\Delta s$ versus density at several representative temperatures for (a) the triangle-well potential and (b) the ramp potential, in both cases with $\lambda=1.5$. }
	\label{fig:RMPE}
\end{figure}

\subsection{Residual Multiparticle Entropy}
It is well known that  the excess entropy per particle $s_\ex$ [see Eq.\ \eqref{sex}] can be expressed as an infinite sum of contributions associated with spatially integrated density correlations of increasing order \cite{BE89,GG92,NG58}. In the absence of external fields, the leading and quantitatively dominant term of the series is the so-called ``pair~entropy'' given by Eq.\ \eqref{2.9a}. The net contribution to $s_\ex$ due to spatial correlations involving three, four, or more particles is represented by the RMPE $\Delta s$.

The density dependence of the RMPE is shown in Fig.\ \ref{fig:RMPE} for the same representative temperatures as considered before. As expected \cite{G08,GG92,SSG18}, the RMPE starts being negative, reaches a minimum value $\Delta s_{\min}$ at a certain density $n_{\min}^*$, and thereafter it grows rapidly, becoming positive beyond a density $n_0^*>n_{\min}^*$. In the case of the triangle-well potential, the values of $|\Delta s_{\min}|$, $n_{\min}^*$, and $n_0^*$ decrease with decreasing temperature. On the other hand, in the case of the ramp potential, the RMPE is much less sensitive to temperature, especially in what concerns $n_0^*$; apart from that, $|\Delta s_{\min}|$ exhibits a nonmonotonic behavior (first increasing and then decreasing with decreasing temperature), whereas $n_{\min}^*$  decreases as temperature decreases.

\begin{figure}%[htbp]
	\centering
	\subfigure[$n^*=0.6$, Triangle well]{\includegraphics[height=5cm]{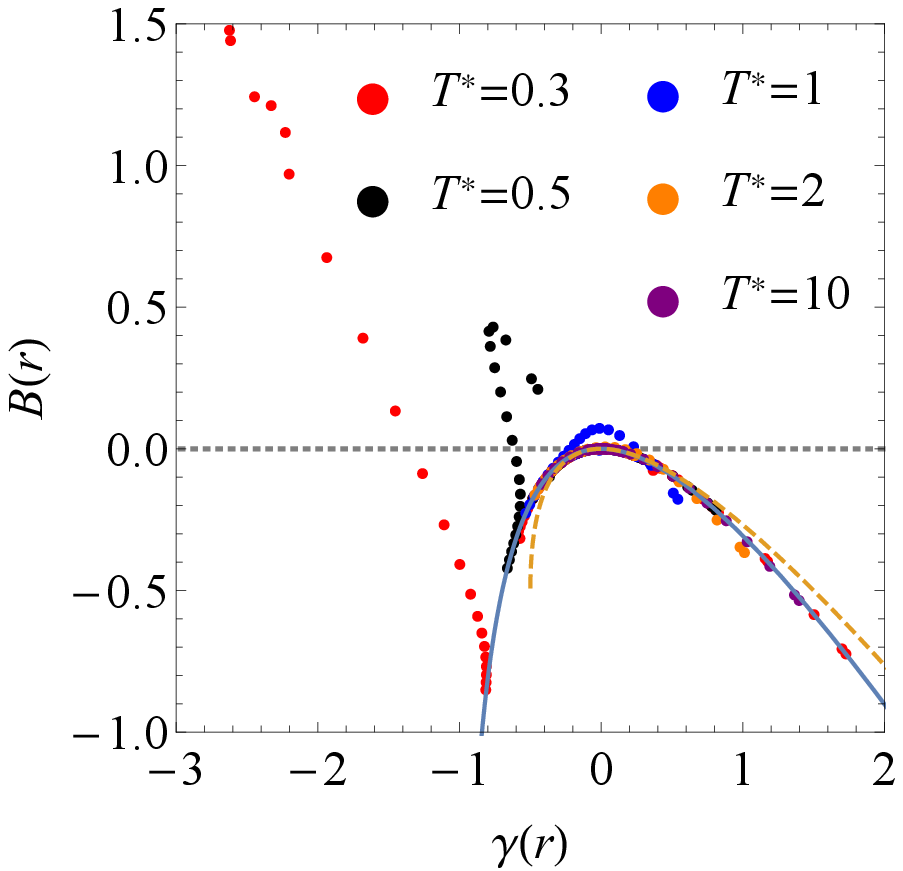} \label{bridge_T_well}}\hspace{1cm}
	\subfigure[$n^*=0.6$, Ramp]{\includegraphics[height=5cm]{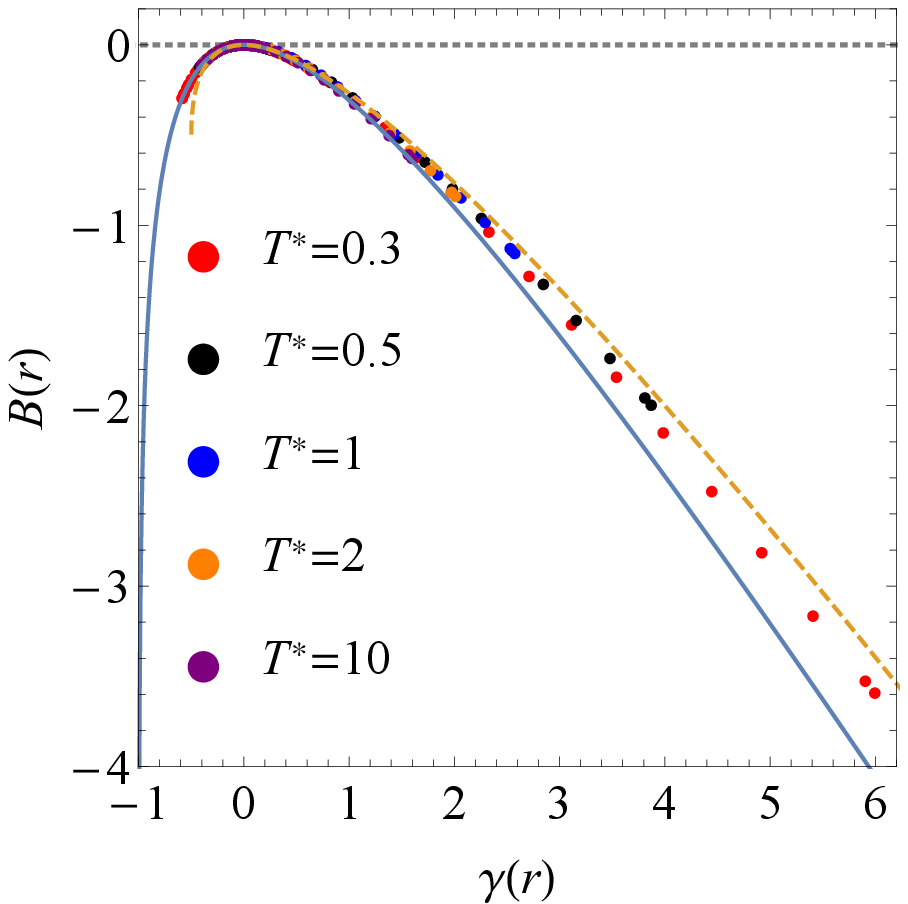} \label{bridge_T_ramp}}\\
\subfigure[$T^*=1$, Triangle well]{\includegraphics[height=5cm]{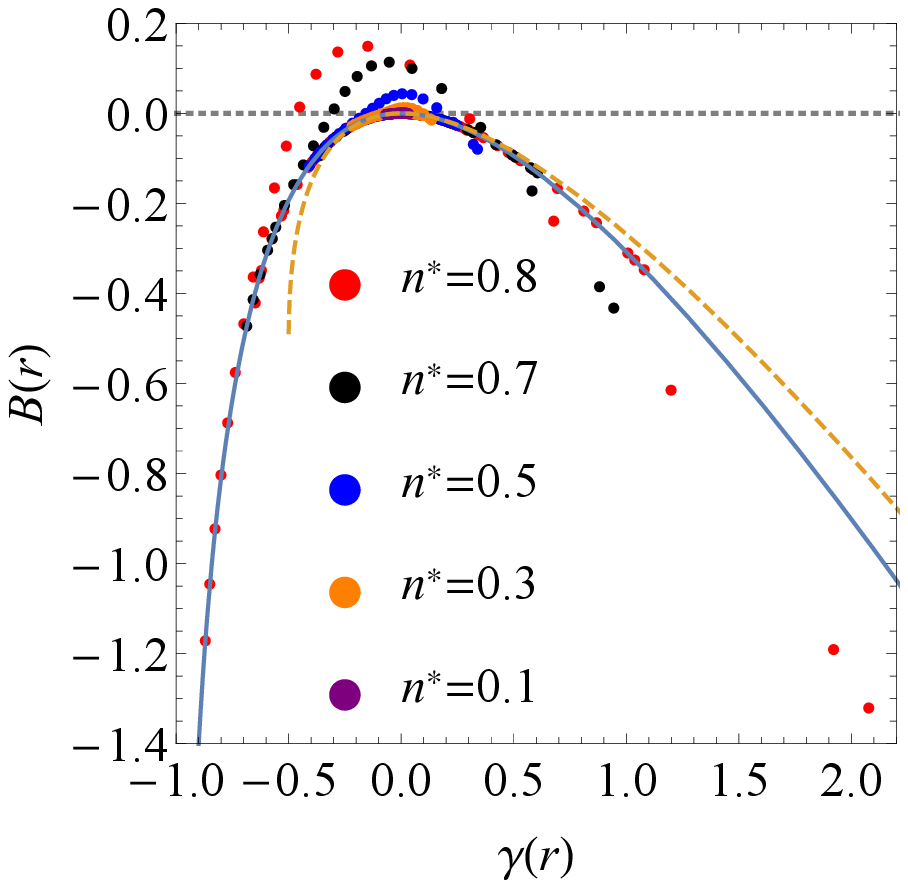} \label{bridge_n_well}}\hspace{1cm}
	\subfigure[$T^*=1$, Ramp]{\includegraphics[height=5cm]{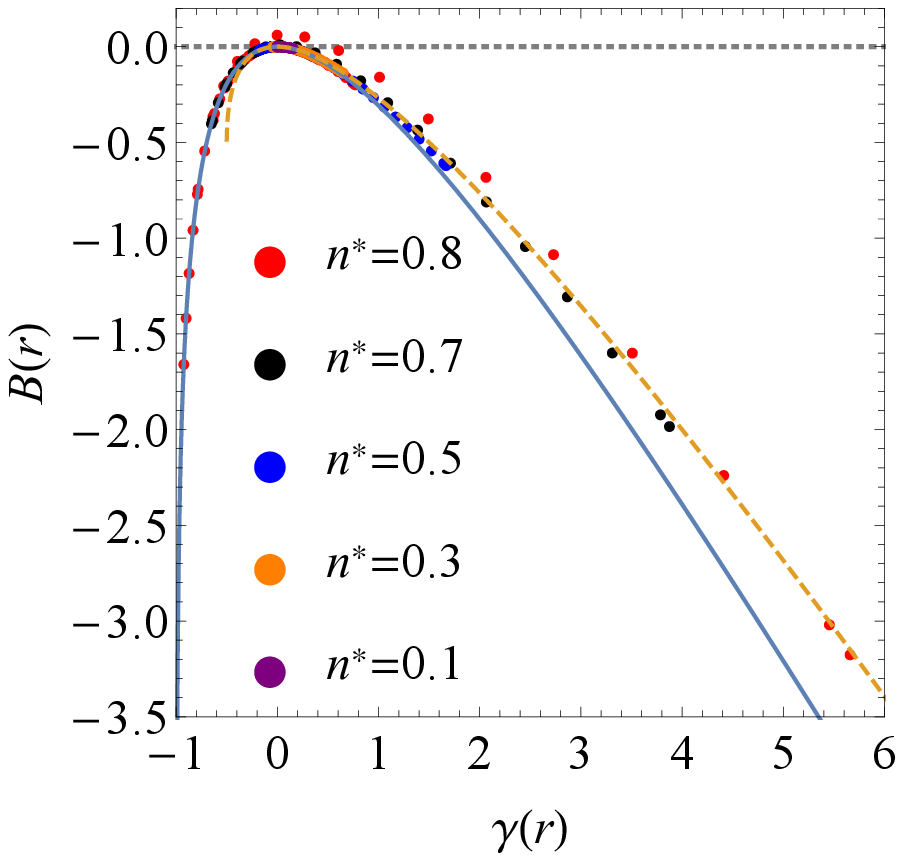} \label{bridge_n_ramp}}
	\caption{Scatter plot of the bridge function $B(r)$ versus the indirect correlation function $\gamma(r)$ at several representative temperatures for a reduced density $n^*=0.6$ [panels (a) and (b)] and at several representative densities for a reduced temperature $T^*=1$ [panels (c) and (d)]. Panels (a) and (c) correspond to the triangle-well potential, while panels (b) and (d) correspond to the ramp potential, in all cases with $\lambda=1.5$. The solid and dashed lines represent the approximate closures $B_{\text{PY}}[\gamma]=\ln\left(1+\gamma\right)-\gamma$ and $B_{\text{MS}}[\gamma]=\sqrt{1+2\gamma}-1-\gamma$, respectively. }
	\label{fig:bridge}
\end{figure}

\section{Assessment of the  Hypernetted-Chain, Percus--Yevick, and Martynov--Sarkisov Closures}
\label{sec4}
In the classical theory of liquids \cite{HM06,S16}, the structural properties are usually obtained by supplementing the Ornstein--Zernike relation $\widetilde{h}(k)=\widetilde{c}(k)+n\widetilde{c}(k)\widetilde{h}(k)$ [see Eq.\ \eqref{2.7}] with an approximate \emph{closure}. Such a closure is frequently expressed as a functional dependence of the so-called bridge function
\beq
\label{eq:bridge}
B(r)\equiv \ln g(r)+\beta\phi(r)-\gamma(r)
\eeq
and the indirect correlation function $\gamma(r)$, i.e., $B(r)\simeq B_{\text{approx}}[\gamma(r)]$. Three of the most popular closures are the hypernetted-chain (HNC) \cite{M58,vLGB59}, Percus--Yevick (PY) \cite{PY58}, and Martynov--Sarkisov (MS) \cite{MS83} ones. They read
\beq
\label{closures}
B_{\text{HNC}}[\gamma]=0,\quad B_{\text{PY}}[\gamma]=\ln\left(1+\gamma\right)-\gamma,\quad B_{\text{MS}}[\gamma]=\sqrt{1+2\gamma}-1-\gamma.
\eeq

In order to asses the performance of those closures, in Fig.\ \ref{fig:bridge} we present scatter plots of $B(r)$ versus $\gamma(r)$, both quantities being obtained from the exact solution. In those plots we have restricted ourselves to the region $r>1$ due to the singular character of both $\ln g(r)$ and $\beta\phi(r)$ in the region $r<1$.

In the case of the triangle-well potential, it is clearly seen that a ``universal'' branch (in the sense that points corresponding to different states collapse into a common curve) coexists with state-dependent branches, the latter being quite apparent at low temperatures [see $T^*=0.3$, $0.5$, and $1$ in Fig.\ \ref{bridge_T_well}] or high densities [see $n^*=0.8$, $0.7$, and $0.5$ in Fig.\ \ref{bridge_n_well}]. This effect is much less noticeable for the ramp potential and in that case it is essentially restricted to high densities [see $n^*=0.8$ in Fig.\ \ref{bridge_n_ramp}]. We have noticed that, at a given state, the nonuniversal branch corresponds to the values of $\gamma(r)$ and $B(r)$ in the first coordination shell (i.e., $1<r<2$). A consequence of the existence of the nonuniversal branch is the lack of one-to-one correspondence between $\gamma$ and $B$. For instance, in the triangle-well fluid with $n^*=0.8$ and $T^*=1$, one has $(\gamma,B)=(-0.378,0.087)$ at $r=1.360$ (nonuniversal branch), while $(\gamma,B)=(-0.378,-0.097)$ at $r=2.748$ (universal branch).

In what concerns the closures in Eq.\ \eqref{closures}, note that the HNC approximation $B_{\text{HNC}}[\gamma]=0$ clearly fails since values $|B|\gtrsim 1$ are reached, especially at low temperatures and/or high densities, in both interaction models. The universal branch is reasonably well captured by both $B_{\text{PY}}[\gamma]$ (which becomes meaningless if $\gamma<-1$) and $B_{\text{MS}}[\gamma]$ (which becomes meaningless if $\gamma<-0.5$). The PY closure turns out to be more accurate than the MS one for triangle-well fluids. On the other hand, in the case of the ramp fluid, the PY closure is preferable for $\gamma<0$, while the opposite happens if $\gamma\gtrsim 2$. Overall, one can conclude that the PY approximation presents the best global performance. This is not unexpected since it is known that the PY closure  provides the exact solution for hard rods \cite{YS93a}.

\begin{figure}%[htbp]
	\centering
	\subfigure[]{\includegraphics[height=5cm]{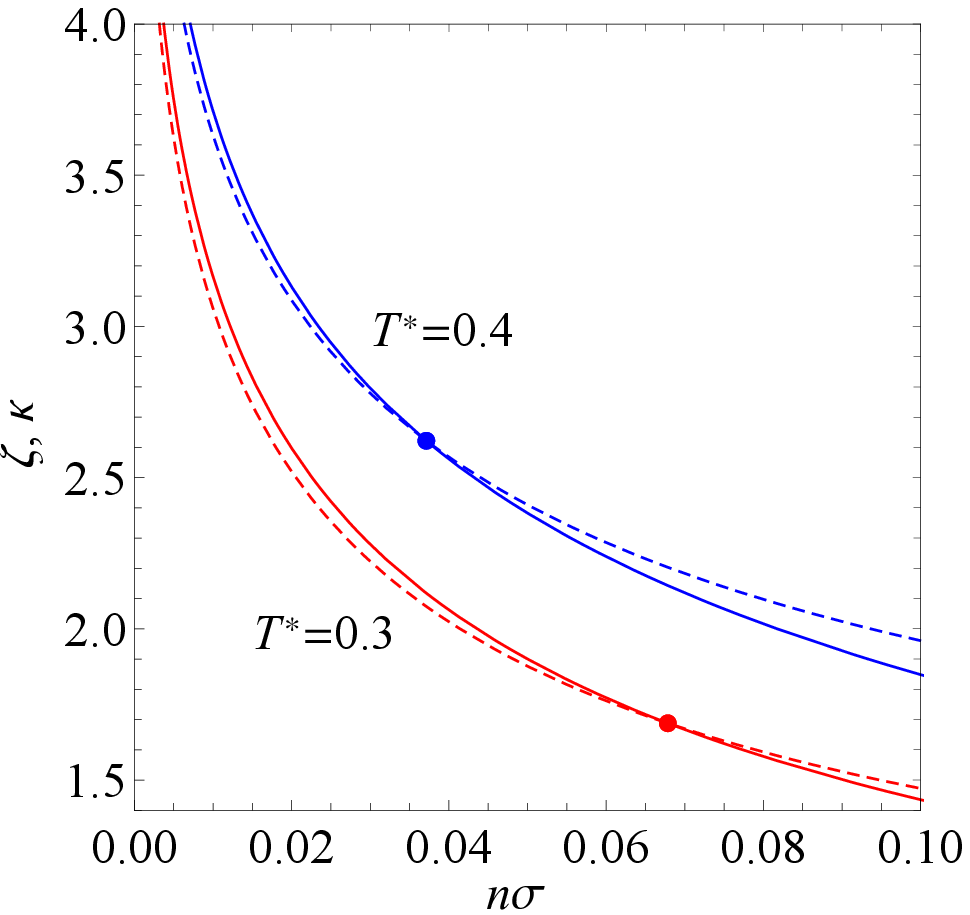} \label{zeta_kappa}}\hspace{1cm}
	\subfigure[]{\includegraphics[height=5cm]{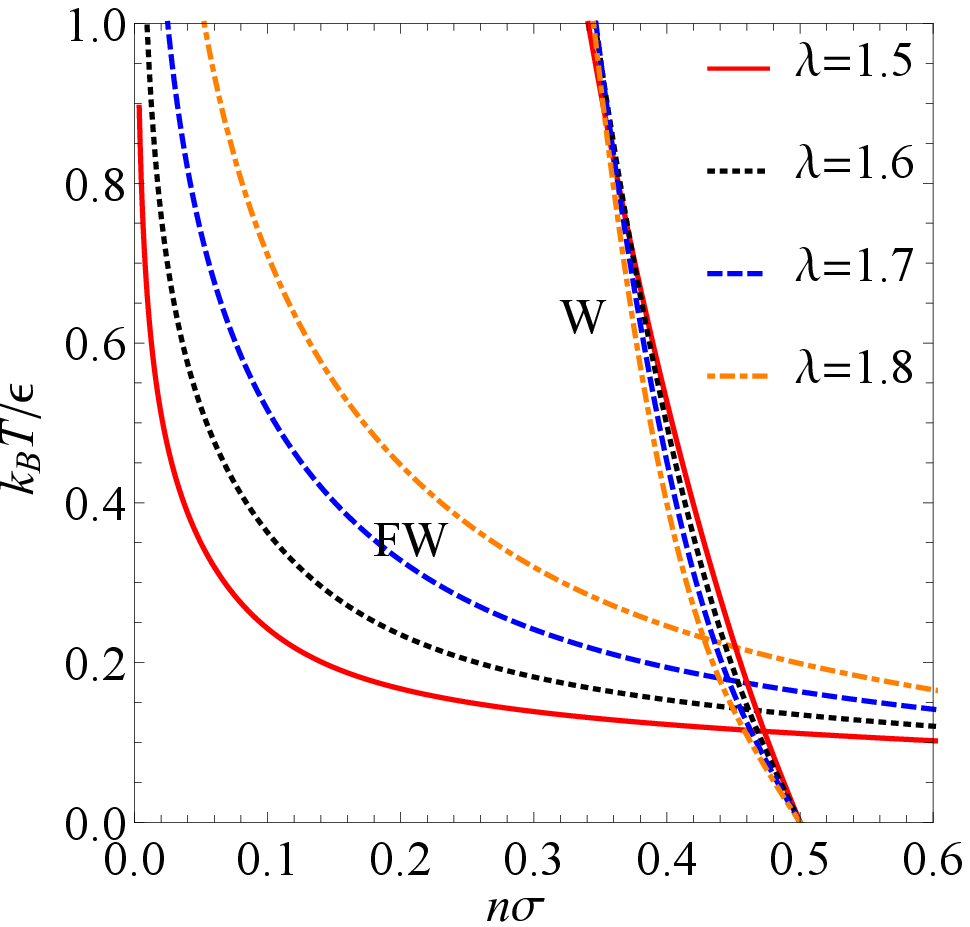} \label{FW_W}}
	\caption{(a) Plot of the damping coefficients $\zeta$ (solid lines) and $\kappa$ (dashed lines) versus density at $T^*=0.3$ and $T^*=0.4$ in the triangle-well fluid with $\lambda=1.5$. At each temperature, the circle marks the intersection point $\gamma=\kappa$ (FW transition point). Although not shown in the graph, the $\kappa$-curves present minima ($\kappa=1.00$ if $T^*=0.3$, $\kappa=1.41$ if $T^*=0.4$) at $n^*=0.44$ and $n^*=0.42$ for $T^*=0.3$ and $T^*=0.4$, respectively; those minima define the Widom line. (b)
Phase diagram in the $T^*$-$n^*$ plane for the triangle-well fluid with different values of the range $\lambda$.
In each case, the FW line splits the diagram into the lower region, where the decay of $h(r)$ is monotonic (i.e., $\kappa<\zeta$), and the upper region, where the decay is oscillatory (i.e., $\zeta<\kappa$). The curves labeled ``W'' represent Widom lines, where, at a given temperature, $\kappa$ presents a minimum value. }
	\label{fig:zeta_kappa_FW_W}
\end{figure}

\section{Fisher--Widom and Widom Lines for the Triangle-Well Potential}
\label{sec5}
While Eq.\ \eqref{g(r)2} provides an exact and fully analytic expression of $g(r)$ up to any finite distance $r$, it does not become practical if one is interested in the asymptotic behavior in the limit $r\to\infty$. Such an asymptotic behavior is directly related to the poles $s_i$ of the Laplace transform $\GG(s)$, i.e., the roots of  $\Omega_\beta(s_i+\beta p)=\Omega_\beta(\beta p)$ [see Eq.\ \eqref{6.6}]. Therefore \cite{DE00,EHHPS93,FW69,PBH17},
\beq
\label{sec:fw1}
h(r)= {A}_1 \rme^{s_1 r}+{A}_2 \rme^{s_2 r}+{A}_3 \rme^{s_3 r}+\cdots,
\eeq
where the  ordering $0>\mathrm{Re}(s_1)\geq \mathrm{Re}(s_2)\geq \mathrm{Re}(s_3)\geq \cdots$ is assumed, and the amplitudes ${A}_i=\mathrm{Res}\left[\GG(s)\right]_{s_i}$ are the associated residues. Thus, the {asymptotic} decay  of  $h(r)$
is determined by the nature of the pole(s)  with the largest real part.

In general, in the case of potentials with an attractive part (as happens in the triangle-well model), the three dominating terms in Eq.\ \eqref{sec:fw1} are those associated with a pair of complex conjugate poles $s_{1,2}=-\zeta\pm\rmi \omega$ (or $s_{2,3}=-\zeta\pm\rmi \omega$) and a real pole $s_3=-\kappa$ (or $s_1=-\kappa$). Therefore, the dominant behavior of $h(r)$ at large $r$ is
\begin{equation}
\label{asympt}
h(r)\approx
\begin{cases}
2 |{A}_\zeta| {\rme^{-\zeta r}}\cos(\omega r+\delta),& \zeta<\kappa,\\
{A}_\kappa {\rme^{-\kappa r}},&\zeta>\kappa,
\end{cases}
\end{equation}
where $\delta$ is the argument of the residue ${A}_{\zeta}$, i.e., ${A}_{\zeta}=|{A}_{\zeta}|\rme^{\pm \rmi\delta}$.
Equation \ref{asympt} shows that the asymptotic behavior of $h(r)$ results from the competition between $\zeta$ and $\kappa$: if $\zeta<\kappa$, a pair of conjugate complex poles dominate and the decay of the total correlation function is \emph{oscillatory}; on the other hand, if $\kappa<\zeta$, a real pole  is the dominant one and then the asymptotic decay is \emph{monotonic},  $\xi=\kappa^{-1}$ representing the \emph{correlation length}.

The oscillatory (monotonic) decay reflects the effects of the repulsive (attractive) part of the interaction potential. Accordingly, at a given temperature, the oscillatory (monotonic) form dominates at sufficiently high (low) values of density.
This is shown in Fig.\ \ref{zeta_kappa}, where the density dependence of $\zeta$ and $\kappa$ is presented at two temperatures for the triangle-well potential with $\lambda=1.5$.
Following Fisher and Widom \cite{FW69}, the locus of transition points from one type of decay to the other one ($\zeta=\kappa$) defines a line, the so-called {Fisher--Widom (FW) line}, in the temperature-versus-density plane \cite{B96,DE00,EHHPS93,HRYS18,TCV03,VRL95,SBHPG16}. The FW line is plotted in Fig.\ \ref{FW_W} for several values of the interaction range $\lambda$. As expected on physical grounds, the region below the FW line (where the decay is monotonic due to the influence of the attractive tail) grows with $\lambda$.

It is interesting to note that, while (at a given temperature $T^*$) $\zeta$ monotonically decreases with increasing density, $\kappa$ presents a nonmonotonic behavior with a minimum value at a certain density [outside the interval shown in Fig.\ \ref{zeta_kappa}]. The loci in the plane $T^*$-$n^*$ where $\kappa$ has a minimum are also included in Fig.\ \ref{FW_W} for the same values of $\lambda$ as in the FW lines. Although the loci  extend to the region of oscillatory decay, they are physically relevant  in the region of monotonic decay (i.e., below the corresponding FW line), where each locus defines a Widom line (marking the states with a maximum correlation length $\xi=\kappa^{-1}$  at a given temperature) \cite{BMAGPSSX09,HRYS18,LXASB15,RBMI17,RDMS17,XBAS06,XEBS06,XKBCPSS05}.

A few comments are in order. First, it can be seen that, for a given value of $\lambda$, the density range supporting the Widom line is much narrower than that of the FW line. Also, the impact of $\lambda$ on the Widom line is very limited, in contrast to what happens with the FW line.
In two- and three-dimensional systems, the Widom line terminates at a critical point $(n_c^*,T_c^*)$, with a nonzero $T_c^*$, where $\xi\to\infty$. Of course, no critical point exists in one-dimensional systems \cite{H63,R99,vH50}, and therefore the Widom lines in Fig.\ \ref{FW_W} extend up to $T^*=0$. The interesting point is that, as observed in Fig.\ \ref{FW_W} and proved in Appendix \ref{appB}, the Widom line intercepts the zero-temperature axis at $n^*=\frac{1}{2}$, regardless of the value of $\lambda$. From that point of view, one can say that a one-dimensional triangle-well fluid has a ``putative'' critical point $(n_c^*,T_c^*)=(\frac{1}{2},0)$.

\section{Conclusions}
\label{sec6}
In this work we have presented a comprehensive study on the thermophysical and structural properties of two prototypical classes of fluids confined in a one-dimensional line. Both potentials have an impenetrable core (of diameter $\sigma=1$) plus a continuous linear part between $r=1$ and $r=\lambda$. While the mathematical form of that additional part is analogous in both cases, the physical meaning is not. In the triangle-well potential the tail is attractive and, apart from its own physical interest, the main importance of the potential resides in  representing the effective colloid-colloid interaction in the Asakura--Oosawa mixture \cite{BE02}. On the other hand, the ramp potential is purely repulsive with a softened core between $r=1$ and $r=\lambda$.

The exact statistical-mechanical solution in the isothermal-isobaric ensemble has been applied to the study of the equation of state, the excess internal energy per particle, the structure factor, and the radial distribution function. In the latter case, a fully analytic representation for $g(r)$ has been derived in terms of a finite number of coordination-shell terms for any finite $r$ [see Eq.\ \eqref{g(r)2}]. From a numerical Fourier inversion, the indirect correlation function $\gamma(r)$ can be easily obtained taking advantage of the fact that it is a continuous function. Then, the direct correlation function is simply obtained as $c(r)=h(r)-\gamma(r)$.

As a bridge between thermodynamic and structural properties, we have studied the RMPE $\Delta s$, i.e., the net contribution to the excess entropy per particle due to spatial correlations involving three, four, or more particles. As expected, $\Delta s$ is negative for low densities but subsequently reaches a minimum and rapidly changes from negative to positive around a certain density.
This sharp crossover from negative to positive values  suggests that at high enough densities multiparticle correlations
play an opposite role with respect to that exhibited in a low packing regime in that they temper the
decrease of the excess entropy that is largely driven by the pair entropy.

The knowledge of both $g(r)$ and $\gamma(r)$ has allowed us to determine the so-called bridge function $B(r)$ [see Eq.\ \eqref{eq:bridge}]. A scatter plot of $B(r)$ versus $\gamma(r)$ has been used to assess the reliability of three standard closures, HNC, PY, and MS. We have observed that none of the closures capture the existence of nonuniversal branches, especially in the triangle-well model. Those branches are due to the values of $\gamma(r)$ and $B(r)$ in the region $1<r<2$. As for the universal branch (where points corresponding to different states overlap), one can conclude that the best overall performance corresponds to the PY closure.

While in the ramp potential, being purely repulsive, the asymptotic decay of $h(r)$ is always oscillatory, a FW transition line exists in the triangle-well fluid separating a repulsive-dominated region (high temperatures or densities) where the decay is oscillatory from an attractive-dominated  region (low temperatures or densities) where the decay is monotonic. We have analyzed the FW line from the poles of the Laplace transform $\GG(s)$, finding that, as expected, the attractive-dominated region grows as the range of the potential increases. Given a temperature below the FW line, the correlation length reaches a maximum value at a certain density, which defines the so-called Widom line. The value of the density on the Widom line depends very weakly on temperature and is hardly dependent on $\lambda$. In fact, in the low-temperature limit the Widom line terminates at $n^*=\frac{1}{2}$ regardless the value of $\lambda$.

Finally, although the exact solution worked out here is constrained to nearest-neighbor interactions (i.e., $\lambda\leq 2$), an approximate description for $\lambda>2$ would still be possible by applying the methods developed in Ref.\ \cite{FS17}.

\begin{acknowledgements}
A.M.M. is grateful to the Ministerio de Educaci\'on, Cultura y Deporte (Spain) for a Beca-Colaboraci\'on during the academic year 2016--2017, which gave rise to this work. The research of A.S. has been supported by the Spanish Agencia Estatal de Investigaci\'on through Grant No.\ FIS2016-76359-P
and the Junta de Extremadura (Spain) through Grant No.\ GR18079, both partially financed by Fondo Europeo de
Desarrollo Regional funds.

\end{acknowledgements}

\appendix

\section{Inverse Laplace Transform of $F_{\ell\ell_1}(s)$}
\label{appA}
Let us start by considering the following mathematical identity,
\beq
\label{A4}
1=x^{\ell_1}\sum_{j=0}^{\ell-1}\binom{\ell_1+j-1}{\ell_1-1}(1-x)^{j}+(1-x)^\ell\sum_{j_1=0}^{\ell_1-1}\binom{\ell+j_1-1}{\ell-1} x^{j_1}, \quad \ell,\ell_1\geq 1,
\eeq
which can be proved by induction. Dividing both sides by $(1-x)^\ell x^{\ell_1}$, Eq.\ \eqref{A4} becomes
\beq
\label{A1}
\frac{1}{(1-x)^\ell x^{\ell_1}}=\sum_{j=1}^{\ell}\frac{\binom{\ell+\ell_1-j-1}{\ell_1-1}}{(1-x)^{j}}+\sum_{j_1=1}^{\ell_1}
\frac{\binom{\ell+\ell_1-j_1-1}{\ell-1}}{x^{j_1}}.
\eeq

Next, the change of variable $s=-a_\beta x-\beta p $ yields
\beq
\label{A15}
F_{\ell\ell_1}(s)=
\sum_{j=1}^{\ell}\frac{\binom{\ell+\ell_1-j-1}{\ell_1-1}}{a_\beta^{\ell+\ell_1-j}}\frac{(-1)^{\ell_1}}{(a_\beta+\beta p+s)^{j}}+\sum_{j_1=1}^{\ell_1}\frac{\binom{\ell+\ell_1-j_1-1}{\ell-1}}{a_\beta ^{\ell+\ell_1-j_1}}\frac{(-1)^{\ell_1-j_1}}{(\beta p+s)^{j_1}},
\eeq
where $F_{\ell\ell_1}(s)$ is defined in Eq.\ \eqref{Fs}.
Making use of the laplace relation $\mathcal{L}^{-1}\{(a+s)^{-\ell}\}=\Theta(r) \rme^{-a r}r^{\ell-1}/(\ell-1)!$, where $\Theta(\cdots)$ is the Heaviside step function, the inverse Laplace transform of $F_{\ell\ell_1}(s)$ is
\bal
\label{A16}
f_{\ell\ell_1}(r)\equiv&\mathcal{L}^{-1}\left\{F_{\ell\ell_1}(s)\right\}\nn
=&\frac{(-1)^{\ell_1}}{a_\beta^{\ell+\ell_1}}\rme^{-\beta pr}\Theta(r)\left[
\rme^{-a_\beta r}
\sum_{j=1}^{\ell}\frac{\binom{\ell+\ell_1-j-1}{\ell_1-1}a_\beta^{j}}{(j-1)!}r^{j-1}+\sum_{j_1=1}^{\ell_1}\frac{\binom{\ell+\ell_1-j_1-1}{\ell-1}
(-a_\beta)^{j_1}}{(j_1-1)!}r^{j_1-1}
\right].
\eal
In the special case $\ell_1=0$, one simply has
\beq
f_{\ell 0}(r)=\rme^{-(a_\beta+\beta p)r} \frac{r^{\ell-1}}{(\ell-1)!}\Theta(r).
\eeq

\section{Widom Line in the Low-Temperature Limit}
\label{appB}
The Widom line is determined by the two conditions
\beq
\Omega_\beta(-\kappa+\beta p)=\Omega_\beta(\beta p),\quad \Omega'_\beta(-\kappa+\beta p)=\Omega'_\beta(\beta p).
\label{B1}
\eeq
The first equality is the condition for $s=-\kappa$ to be a pole of $\GG(s)$, in agreement with Eq.\ \eqref{6.6}. The second equality is the extremum condition $\left(\partial \kappa/\partial {\beta p}\right)_{\beta}=0$.

From an inspection of Eqs.\ \eqref{eq:Omega} and \eqref{eq:Omega'} one can check that, in the low-temperature limit ($\beta\epsilon\gg 1$), the solutions to Eq.\ \eqref{B1} scale as
\beq
\beta p \to \sqrt{x\frac{a_\beta}{\Xbeta}},\quad \kappa\to y \beta p,
\label{B2}
\eeq
where the parameters $x$ and $y$ are pure numbers to be determined.
Insertion of the scaling laws \eqref{B2} yields
\begin{subequations}
\beq
\lim_{\beta\epsilon\to\infty} \beta p\left[\Omega_\beta(-\kappa+\beta p)-\Omega_\beta(\beta p)\right]=y\left(x-\frac{1}{y-1}\right),
\eeq
\beq
\lim_{\beta\epsilon\to\infty} (\beta p)^2\left[\Omega_\beta'(-\kappa+\beta p)-\Omega_\beta'(\beta p)\right]=1-\frac{1}{(y-1)^2}.
\eeq
\end{subequations}
Then, the conditions \eqref{B1} imply $x=1$, $y=2$.

Next, inserting $\beta p\to \sqrt{a_\beta/\Xbeta}$ into Eq.\ \eqref{eq:density-tw} and taking the limit $\beta\epsilon\to\infty$, one obtains $n^*\to \frac{1}{2}$. It is worth noticing that in the case of a one-dimensional square-well fluid of range $\lambda$, the low-temperature limit of the Widom line is described by $\beta p \to \sqrt{2/(\lambda^2-1)\Xbeta}$, $\kappa\to 2 \beta p$, and $n^*\to 1/(\lambda+1)$.

%\bibliographystyle{spbasic}      % basic style, author-year citations *DO NOT USE!*

    \bibliographystyle{spmpsci}      % mathematics and physical sciences *FOR FINAL VERSION*
%\bibliographystyle{spphys}       % APS-like style for physics *FOR WORKING VERSIONS*


    %\bibliography{D:/Dropbox/Mis_Dropcumentos/bib_files/Liquid}

    \end{document}